\documentclass[aps,prx,floatfix,nofootinbib,onecolumn,superscriptaddress,longbibliography]{revtex4-2}
\usepackage{dsfont}

\usepackage{braket}
\usepackage{siunitx}
\usepackage{multirow}
\usepackage{array}
\usepackage{dcolumn}

\newcolumntype{d}{D{.}{,}{-1}}
\usepackage{amsthm}
\usepackage{amsmath}
\usepackage{amsfonts}
\usepackage{mathtools}
\usepackage{physics}
\usepackage{amssymb}
\usepackage{graphicx}
\usepackage{color}
\usepackage{accents}
\usepackage{bbold}
\usepackage[colorlinks=true, citecolor=blue]{hyperref}
\usepackage[normalem]{ulem}
\usepackage[shortlabels]{enumitem}
\usepackage{float}
\usepackage{caption}
\usepackage{subcaption}
\usepackage{siunitx}
\usepackage[version=3]{mhchem}
\usepackage{booktabs}
\usepackage{todonotes}

\makeatletter 
\newsavebox{\@brx}
\newcommand{\llangle}[1][]{\savebox{\@brx}{\(\m@th{#1\langle}\)}%
  \mathopen{\copy\@brx\kern-0.5\wd\@brx\usebox{\@brx}}}
\newcommand{\rrangle}[1][]{\savebox{\@brx}{\(\m@th{#1\rangle}\)}%
  \mathclose{\copy\@brx\kern-0.5\wd\@brx\usebox{\@brx}}}
\makeatother

\theoremstyle{definition}

\theoremstyle{remark}

\usepackage{listings} 
\usepackage{xcolor}

\definecolor{codegreen}{rgb}{0,0.6,0}
\definecolor{codegray}{rgb}{0.5,0.5,0.5}
\definecolor{codepurple}{rgb}{0.58,0,0.82}
\definecolor{backcolour}{rgb}{0.95,0.95,0.92}
\lstdefinestyle{mystyle}{
  backgroundcolor=\color{backcolour},   commentstyle=\color{codegreen},
  keywordstyle=\color{magenta},
  numberstyle=\tiny\color{codegray},
  stringstyle=\color{codepurple},
  basicstyle=\ttfamily\scriptsize,
  breakatwhitespace=false,         
  breaklines=true,                 
  captionpos=b,                    
  keepspaces=true,                 
  showspaces=false,                
  showstringspaces=false,
  showtabs=false,
  xleftmargin=0.02\textwidth,
  rulecolor=\color[RGB]{200,200,200},
  frame=bt,
  framextopmargin=2pt,
  framexbottommargin=2pt,
  framexleftmargin=10pt,
  tabsize=2
}
\long\def\/*#1*/{}

\begin{document}

\author{Nicholas C. Rubin}
\email{nickrubin@google.com}
\affiliation{
             Google Quantum AI, 
             San Francisco, CA, USA}

\author{Guang Hao Low}
\email{guanghaolow@google.com}
\affiliation{
             Google Quantum AI, 
             San Francisco, CA, USA}

\author{A. Eugene DePrince III}
\email{adeprince@fsu.edu}
\affiliation{
             Department of Chemistry and Biochemistry,
             Florida State University,
             Tallahassee, FL 32306-4390, USA}
\affiliation{
             Google Quantum AI, 
             San Francisco, CA, USA}

\title{Near-frustration-free electronic structure Hamiltonian representations and lower bound certificates}
\begin{abstract}
Hamiltonian representations based on the sum-of-squares (SOS) hierarchy provide rigorous lower bounds on ground-state energies and facilitate the design of efficient classical and quantum simulation algorithms. This work presents a unified framework connecting SOS decompositions with variational two-particle reduced density matrix (v2RDM) theory. We demonstrate that the ``weighted'' SOS ansatz naturally recovers the dual of the v2RDM program, enabling the strict enforcement of symmetry constraints such as particle number and spin. We provide explicit SOS constructions for the Hubbard model and electronic structure Hamiltonians, ranging from spin-free approximations to full rank-2 expansions. We also highlight theoretical connections to block-invariant symmetry shifts. Numerical benchmarks on molecular systems and Iron-Sulfur clusters validate these near frustration-free representations, demonstrating their utility in improving spectral gap amplification and reducing block encoding costs in quantum algorithms.
\end{abstract}

\maketitle
\section{Introduction}
Optimizing a Hamiltonian's representation is essential for classical and quantum simulation, particularly to enhance the scaling of approximate methods. In the context of electronic structure, utilizing Hamiltonian structure has consistently improved constant factors and scaling—for example, by exploiting the low-rank properties of the Coulomb kernel~\cite{werner2003fast}, spatial- or spin-symmetries~\cite{helgaker2013molecular}, tensor-hypercontraction~\cite{hohenstein2012tensor, PRXQuantum.2.030305}, or matrix-product operator representations~\cite{chan2016matrix}. In this work, we add to this set by describing a protocol to obtain a representation of the electronic-structure Hamiltonian that can be used to incorporate low-energy simulation assumptions into classical and quantum algorithms and provides a variational lower bound certificate to the ground-state energy. Specifically, we focus on the working equations for deriving this form of the electronic structure Hamiltonian and the connection between our protocol and its dual, which is known as the pseudo-moment, quantum marginal optimization, or the variational two-particle reduced density matrix (v2RDM) method~\cite{Mazziotti02_062511,Zhao07_553, Zhao08_164113,eugene2024variational}.  

The protocol relies on the fact that any Hermitian operator in a finite basis set that is described using a non-commutative algebra can be expressed as a Hermitian sum of squares (SOS) plus a constant shift ($E_{\text{SOS}}$)~\cite{helton2002positive}
\begin{align}\label{eq:sos_ham_equation}
H - E_{\text{SOS}}\mathbb{1} = \sum_{\alpha}O_{\alpha}^{\dagger}O_{\alpha}.
\end{align}
This equality establishes a non-negative representation of the Hamiltonian and certifies a lower bound on the ground-state energy. The choice of $O_{\alpha}$ that algebraically recovers the Hamiltonian is not unique, and a hierarchy of operators can be obtained by expanding their degree or spatial extent to generate increasingly tight lower bounds. One hierarchy is formed by allowing the degree of the $O_{\alpha}$ polynomial to increase from the minimal degree required to represent the Hamiltonian to the number of particles in the simulation. This method of improving the SOS lower bound recovers the $p$-positivity hierarchy from v2RDM theory \cite{Erdahl01_042113, Fujisawa01_8282, pironio2010convergent}. Another hierarchy that converges $E_\text{SOS}$ to the ground-state energy is obtained by expanding the locality of $O_{\alpha}$, as recently described by Lindsey and Lin~\cite{lin2022variational}. Within the dual sum-of-squares protocol, both hierarchies are unified into one framework which allows for a mixture of $p$-positivity and locality considerations in the generators for the sum-of-squares representation. As an example, we provide a spatially resolved representation of fermionic lattice Hamiltonians that is near frustration free.

In this work, we also highlight that the SOS ansatz alone is insufficient for constructing a set of witness Hamiltonians when considering symmetries such as a fixed particle manifold. 
More generally, the positive ansatz known as the `weighted sum-of-squares', introduced by Helton and McCullough in Ref.~\cite{helton2004positivstellensatz}, can be applied within non-commutative polynomial optimization methods for general quantum Hamiltonians~\cite{pironio2010convergent}.
Recently, Ref.~\cite{PhysRevX.14.031006} showed that one can obtain variational upper and lower bounds on observables other than the energy which vastly expands the utility of these methods. We add to these works by explicit construction of spin-adapted SOS programs, demonstrating how block-invariant symmetry shifts (BLISS)~\cite{loaiza2023block} emerge naturally from a weighted SOS positive ansatz, and illuminate potential applications of SOS representations in both classical and quantum algorithms.   
We also demonstrate that by taking the dual of the v2RDM program in standard form a weighted SOS lower bound certificate is recovered. The SOS construction and mathematical derivation thereof are distinct from prior work, such as that in Ref.~\cite{cances2006electronic}, as well as other similar programs derived by leveraging the Hellmann-Feynman theorem~\cite{PhysRevA.102.052819}.  

While providing lower-bound ground-state energy estimates is an invaluable convergence criterion when combined with variational upper bounds on the ground-state energy, the resulting near-frustration-free Hamiltonian representation can also be used to improve approximate simulation methods. Recent work highlights this strategy in quantum algorithms for ground-state energy estimation, expectation value estimation, and low-energy time evolution~\cite{king2025quantum, low2025fast}. It has also been conjectured that the sign problem in free-projection auxiliary field quantum Monte Carlo can be minimized by utilizing a near frustration free Hamiltonian representation~\cite{hastings2023field}. This work provides the working equations to test these conjectures by providing the mathematical programs and codes necessary to solve for the SOS Hamiltonian representation using a set of spin-adapted generating algebras for different approximations to the quadratic generating algebra.

We structure this work by first introducing the general theory of weighted sum-of-squares and its relationship to BLISS~\cite{loaiza2023block} and the known set of $n$-representability conditions. Second, we provide SOS representations of a Hubbard model at different fillings and demonstrate that the SOS framework allows one to construct $k$-local non-negative representations of Hamiltonians that are near-frustration free. This example demonstrates the dual of Ref.~\cite{lin2022variational}. Finally, we introduce the spin-free generating algebra SOS and the spin-adapted quadratic generating algebra SOS and show how the weighted sum-of-squares recovers lower bounds to the desired particle sector of Fock space for the electronic structure Hamiltonian represented in second quantization.

\section{The SOS Dual}
We first set notation for the mathematical description of the primal and dual certificates.  Our focus will be on Hamiltonians represented in finite non-commutative algebra of fermionic ladder operators. The fermionic ladder operators act on a vector space size $2^{m}$ where $m$ is the number of spin-orbitals. The full vector space is a Fock space composed of a direct sum of fixed $\eta$-particle Hilbert spaces. Each Hilbert space corresponds to the Fock basis states with a fixed number of particles.
\subsection{positivity ansatz}
The variational $2$-RDM method is usually derived by taking the approximation that the dual cone is a sum of squares in ladder operators of degree-$2d$~\cite{Erdahl78_697, Mazziotti04_213001}. This assumption is then translated into a non-negativity constraint on the matrix of pseudoexpectation values which are used as the primal variables in the variational $2$-RDM semidefinite program. Specifically, the dual cone $\mathcal{B}_{\text{SOS}}$ is approximated to be a positive sum of squares 
\begin{align}
\mathcal{B}_{\text{SOS}} \coloneqq \{ B = \sum_{i}R_{i}^{\dagger}R_{i} \;\vert \; R_{i} \in \text{Span}\left( \{a_{k}, a_{l}^{\dagger}, \mathbb{1}\}^{d}\right)\}
\end{align}
where $R_{i}$ is a degree $d$ polynomial of ladder operators with complex coefficients. 
While this structure is sufficient to represent positive polynomials of non-commuting variables~\cite{helton2002positive}, it does not naturally include algebraic constraints or symmetry ({\em e.g.}, particle number, spin, etc.) constraints.  Accounting for such constraints requires a generalized positive ansatz, which is known as a weighted sum of squares~\cite{helton2004positivstellensatz}. The weighted sum of squares features the $\mathcal{B}_{\text{SOS}}$ along with a congruence transformed set of constraints $q_{i} \geq 0$ 
\begin{align}
\mathcal{B}_{\text{wSOS}} \coloneqq \Bigg \{ B &= \sum_{i}R_{i}^{\dagger}R_{i} + \sum_{ji}x_{ij}^{\dagger}q_{j}x_{ij}\; \Bigg \vert \; R_{i} \in \text{Span}\left( \{a_{k}, a_{l}^{\dagger}, \mathbb{1}\}^{d}\right), \nonumber \\
& x_{ij} \in  \text{Span}\left( \{a_{k}, a_{l}^{\dagger}, \mathbb{1}\}^{d'}\right), d' = d - \left\lceil\max(\text{deg}(q_{j}))/2 \right\rceil \Bigg \}
\end{align}
where the polynomials of non-commutative operators $q_{i}$ define the positivity domain of interest. Of particular interest to the quantum chemistry community is the set of linear constraints defining the particle number ($\eta$) manifold of interest, $q = \hat{n} - \eta \mathbb{1} = 0$, or any other symmetry constraint such as spin or spatial symmetry. The anticommutation relations can also be included as constraints for the positivity domain--e.g. $\left\{a_{i}, a_{j}\right\} = 0$.  In these cases, as explained in Ref.~\cite{pironio2010convergent}, it is simpler to include equality constraints as linear constraints. Taking the dual of the primal problem, one can show that the SOS certificate for the equality constraints, $r_{i}$, are of the form
\begin{align}
\sum_{i}f_{i}r_{i} + r_{i}^{\dagger}f_{i}^{\dagger}
\end{align}
where $f_{i}$ are degree $d'$ such that $\text{deg}(f_{i}r_{i}) \leq 2d$.  This is precisely the form of the block-invariant symmetry shift~\cite{loaiza2023block} used to reduce the norm of the Hamiltonian when simulating in a fixed symmetry manifold. Thus, the weighted SOS for the $n$-representability problem is
\begin{align}\label{eq:wsos_nrep}
\mathcal{B}_{\text{wSOS-nrep}}  \coloneqq \Bigg \{ B &= \sum_{i}R_{i}^{\dagger}R_{i} + \sum_{i}f_{i}r_{i} + r_{i}^{\dagger}f_{i}^{\dagger}\; \Bigg \vert \; R_{i} \in \text{Span}\left( \left\{a_{k}, a_{l}^{\dagger}, \mathbb{1}\right\}^{d}\right),  \nonumber \\
 &f_{i} \in  \text{Span}\left( \{a_{k}, a_{l}^{\dagger}, \mathbb{1}\}^{d'}\right), d' = 2d - \max(\text{deg}(r_{i}))\Bigg \} 
\end{align}
which constitutes the complete non-negative ansatz for a non-negative operator restricted to a particular symmetry subspace. Due to the extra non-negativity, or congruency constraint, the weighted SOS enjoys the inclusion relation
\begin{align}\label{eq:cone_subset}
\mathcal{B}_{\text{SOS}} \subseteq \mathcal{B}_{\text{wSOS}}
\end{align}
which indicates that any solution to $n$-representability needs to articulate a dual cone over $\mathcal{B}_{\text{wSOS}}$. Later we demonstrate by direct calculation that the linear constraints imposed in the v2RDM problem lift from $\mathcal{B}_{\text{SOS}}$ to $\mathcal{B}_{\text{wSOS}}$.
\subsection{The Semidefinite program to find the SOS dual}
Given a guess for the operator form of a non-negativity ansatz, one can construct a semidefinite program to determine the coefficients of the SOS generators and thus the SOS representation of the Hamiltonian. Given a set of operators $\{\mathfrak{o}_{j}\}$ any $O_{\alpha}$ is constructed by
\begin{align}
O_{\alpha} = \sum_{j}c_{j}\mathfrak{o}_{j}
\end{align}
where $c_{j} \in \mathbb{C}$. Given $L$ $\mathfrak{o}_{j}$ generators a non-negative operator that is the sum of at most $L$ non-negative  operators can be represented using the Gram matrix $G$
\begin{align}
\sum_{\alpha=1}^{L}O_{\alpha}O_{\alpha} =& \left(\sum_{j'}c_{j'}^{*}\mathfrak{o}_{j}^{\dagger}\right)\left(\sum_{j}c_{j}\mathfrak{o}_{j}\right) \\
=& \Vec{\mathfrak{o}^{\dagger}}\left(\sum_{jj'}c_{j'}^{*}c_{j} \right) \Vec{\mathfrak{o}} = \Vec{\mathfrak{o}^{\dagger}}G \Vec{\mathfrak{o}}.
\end{align}
In order to relate this operator product to the Hamiltonian plus a shift (Eq~\eqref{eq:sos_ham_equation}), pair products of algebra elements $\mathfrak{o}_{j}^{\dagger}\mathfrak{o}_{i}$ are placed in a normal order. A given Hamiltonian expressed as a subset of $\{\mathfrak{o}_{j}\}$ elements can then be used to construct the appropriate equality constraints on $G$ to ensure that, while maximizing the coefficients associated with the identity element of the algebra, the Hamiltonian is recovered. The linear constraints, the positive semidefinite constraint on $G$, and the linear objective associated with the sum of coefficients of the identity algebra elements is a semidefinite program
\begin{align}
\max E_{\text{SOS}}& \\
\text{s.t.} \;& H - E_{\text{SOS}}\mathbb{1} = \Vec{\mathfrak{o}^{\dagger}}G \Vec{\mathfrak{o}} \\
G & \succeq 0.
\end{align}
We have retained the operator notation to emphasize that the linear constraints are generated from normal ordering pair products of the generating algebra, but it should be read that the PSD variables are size $L \times L$--the size of the SOS generating algebra.  If $\mathfrak{o}_{i}^{\dagger}\mathfrak{o}_{j}$ produces terms that do not appear in the Hamiltonian, then a linear constraint must ensure that these terms evaluate to zero. In the following section, we will provide specific examples of the equality constraints for electronic structure relevant algebras.
\section{Sum of Squares Examples}
\subsection{Hubbard Hamiltonian}
In this section we present two representations of the one-dimensional Hubbard Hamiltonian as a sum of squared terms.  The one-dimensional Hubbard Hamiltonian is 
\begin{align}
H = -t \sum_{\langle i, j\rangle, \sigma}a_{i\sigma}^{\dagger}a_{j\sigma} + U \sum_{i}n_{i\alpha}n_{i\beta}
\end{align}
where $\langle i,j\rangle$ denotes indices $i,j$ that are geometrically nearest-neighbor sites on a lattice, and $\sigma \in \{\alpha, \beta\}$ denotes the spin index of the electrons. We consider the SOS generating algebra
\begin{align}\label{eq:sos_fock_nn_algebra}
O_{ij} \in \text{span}\left(\left(\prod_{\sigma\in \{\alpha,\beta\}}\left(\mathbb{1}, a_{i\sigma}, a_{i\sigma}^{\dagger},  n_{i,\sigma}\right)\right) \otimes \left(\prod_{\sigma \in \{\alpha, \beta\}} \left(\mathbb{1}, a_{j\sigma}, a_{j\sigma}^{\dagger},  n_{j,\sigma}\right) \right) \right)
\end{align}
and consider site indices $i, j$ of nearest neighbors. The product, $\prod_{\sigma\in \{\alpha,\beta\}}\left(\mathbb{1}, a_{i\sigma}, a_{i\sigma}^{\dagger},  n_{i,\sigma}\right)$, is shorthand for a cartesian product over spin variables of the set of operators in parenthesis.  The coefficients of each $O_{ij}^{\dagger}O_{ij}$ contribute to a Gram matrix $\boldsymbol{G}_{ij}$ corresponding to the variables used to optimize a local sum-of-squares representation. Using OpenFermion~\cite{mcclean2020openfermion}, we can automatically produce products of algebra elements, normal order them, and then determine if the corresponding Gram matrix element should be counted in the cost function \textcolor{black}{for the semidefinite program (SDP)} ({\em i.e.}, $E_\text{SOS}$), constrained to be equal to a Hamiltonian coefficient (in the case the normal ordered term produces a Hamiltonian term), or constrained to be zero. 

To demonstrate the lower bound quality difference between different algebras we also consider a more restrictive SOS generating algebra that partitions the generators into single ladder operators and the set $\{\mathbb{1}, n_{i\sigma}\}$. Considering the set of SOS generators that produce quadratic operators, the elements of the Gram matrix will correspond to non-negative operators of the form 
\begin{align}
\sum_{\gamma}O_{\gamma}^{\dagger}O_{\gamma} = \sum_\gamma \sum_{\sigma\tau}\sum_{ij} \left ( d_{i\sigma}^\gamma d_{j\tau}^\gamma a^\dagger_{i\sigma} a_{j\tau} + d_{i\sigma}^\gamma q_{j\tau}^\gamma a^\dagger_{i\sigma} a^\dagger_{j\tau} + q_{i\sigma}^\gamma d_{j\tau}^\gamma a_{i\sigma} a_{j\tau} + q_{i\sigma}^\gamma q_{j\tau}^\gamma a_{i\sigma} a^\dagger_{j\tau} \right ) 
\end{align}
where $d$ and $q$ are free parameters to optimize. We further simplify this SOS by separating the particle and hole sectors as
\begin{align}
\hat{H}_{\rm SOS} = \sum_{\gamma} \sum_\sigma O^\dagger_{d_\sigma,\gamma} O_{d_\sigma,\gamma} +\sum_{\gamma} \sum_\sigma O^\dagger_{q_\sigma,\gamma} O_{q_\sigma,\gamma} 
\end{align}
where
\begin{align}
O_{d_\sigma,\gamma} &= \sum_{i} d_{i\sigma}^\gamma a_{i\sigma}  \label{eq:O_ladder_D}\\
O_{q_\sigma,\gamma} &= \sum_i q_{i\sigma}^\gamma a_i^\dagger.\label{eq:O_ladder_Q}
\end{align}
Following the prior nearest-neighbor SOS generating algebra we can restrict the summation indices in Eq~\eqref{eq:O_ladder_D} and Eq.~\eqref{eq:O_ladder_Q} to run over spatial indices of any subset of lattice sites--\textit{i.e.} nearest-neighbor pairs, nearest-neighbor triplets, etc. The resulting Gram matrices have the form
\begin{align}\label{eq:sos_op_delta}
{\bf B}_{\delta} = \begin{pmatrix}
{\bf D}_\alpha(\delta) & 0 & 0 & 0 \\
0 & {\bf D}_\beta(\delta) & 0 & 0 \\
0 & 0 & {\bf Q}_\alpha(\delta) & 0 \\
0 & 0 & 0 & {\bf Q}_\beta(\delta) \\
\end{pmatrix}
\end{align}
where the $\delta$ labels the subset of the lattice sites considered in sums of Eq.~\eqref{eq:O_ladder_D} and Eq.~\eqref{eq:O_ladder_Q}.

There are a variety of ways to convert the Hubbard interaction into an SOS form. Here, we use the site-local generators $O_{\alpha} \in \text{Span}\left( \mathbb{1}, n_{i\sigma} \right)$ resulting in $3\times 3$ Gram matrices 
\begin{align}\label{eq:fock_sos_nn_dec_algbra_two_body}
{\bf B}_{i} = \begin{pmatrix}
b_{11} \mathbb{1} & b_{12} n_{i\alpha} & b_{13} n_{i\beta} \\
b_{21} n_{i\alpha} & b_{22} n_{i\alpha} & b_{23} n_{i\alpha}n_{i\beta} \\
b_{31} n_{i\beta} & b_{32}n_{i\beta}n_{i\alpha} & b_{33} n_{i\beta}
\end{pmatrix}
\end{align}
which are constrained to be positive semidefinite. In Eq.~\eqref{eq:fock_sos_nn_dec_algbra_two_body} we provided the fermionic operators in specific positions of the matrix as a guide to the reader to observe which coefficients of the matrix ${\bf B}_{i}$ correspond to which operators. The operators themselves are not part of the semidefinite program. When computing SOS form of an $L$-site one-dimensional periodic boundary Hubbard model, there are $L$ nearest-neighbor gram matrices ${\bf B}_{ij}$, each containing four blocks, and $L$ onsite gram matrices ${\bf B}_{i}$ in the semidefinite program. The spin-decoupled and spinful algebras just described are examples of an approximate dual cone $\mathcal{B}_{\text{SOS}}$.

In order to find lower bounds to a fixed particle sector of Fock space we construct the weighted SOS non-negative ansatz, $\mathcal{B}_{\text{wSOS-nrep}}$, using the particle number constraint polynomial
\begin{align}
r = \left(\sum_{i}\left(n_{i\alpha} + n_{i\beta}\right)\right) -  \eta \mathbb{1}
\end{align}
which is a degree-2 polynomial of ladder operators. Left and right multiplying by $f$,
\begin{align}\label{eq:r_ideal}
f \in \text{Span}\left( \mathbb{1}, n_{j\sigma} \right),
\end{align}
constructs the weighted SOS Ideal term
\begin{align}\label{eq:weight_in_sos}
r f + f^{\dagger}r =& 2 \left(c \mathbb{1} + \sum_{j,\sigma}d_{j, \sigma}n_{j\sigma}\right)\left(\sum_{i}\left(n_{i\alpha} + n_{i\beta}\right) -  \eta \mathbb{1}\right). 
\end{align}
The weighting term is added as an additional term in the SDP. After normal ordering Eq.~\eqref{eq:weight_in_sos} the variables (coefficients) of $f$ are added to the appropriate linear constraints or added to the cost function.

\begin{figure}[H]
    \centering
    \includegraphics[width=8.5cm]{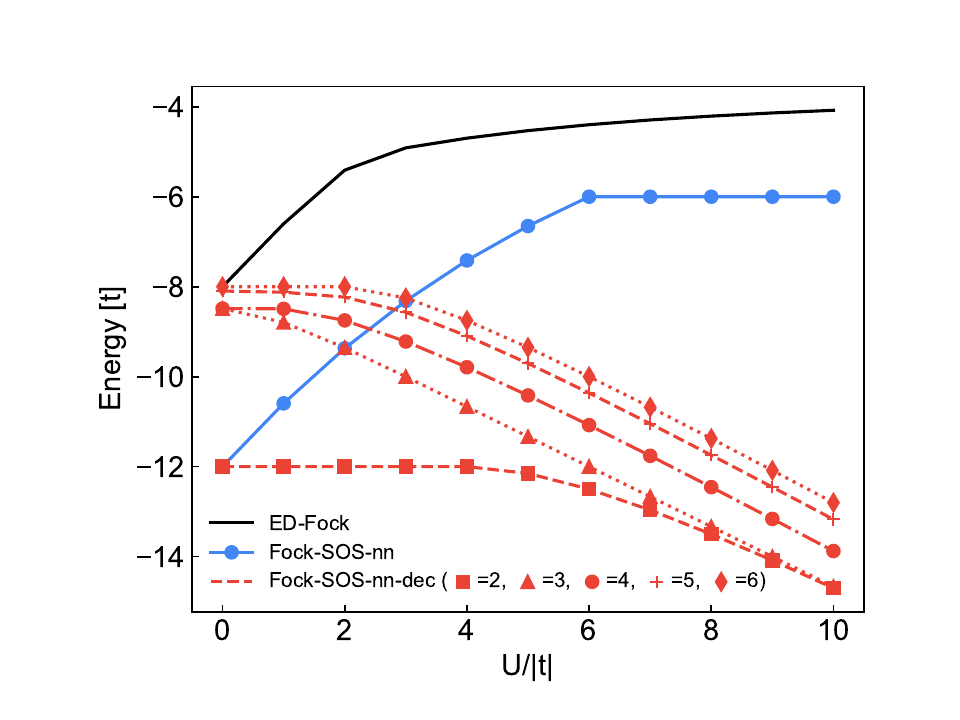}
    \caption{Fock space ground-state energy of the six-site one-dimensional PBC Hubbard model (black) along with energy lower bound certificates (blue and red) computed with the Fock-SOS-nn algebra (blue) and the Fock-SOS-nn-dec algebra (red). The different red curves are labeled by a number which is the largest number of contiguous sites to include in the sum of SOS generators ($\delta$).  \label{fig:fock_space_hub}}
\end{figure}

In Fig.~\ref{fig:fock_space_hub} we show the lower bound obtained from the 6-site one-dimensional periodic boundary Hubbard model using the two aforementioned algebras without a weighted SOS constraint. We used RRSDP~\cite{Burer03_329} implemented in \texttt{libSDP}~\cite{libsdp} to solve the SOS SDP. The SDP stopping criteria was set to  $1\times 10^{-6}$ in the linear constraint residual and the change in the objective function between iterations. In addition, we restricted ourselves to real-valued SDPs. The algebra in Eq.~\eqref{eq:sos_fock_nn_algebra} with no spin decoupling is denoted Fock-SOS-nn and the algebra with spin-decoupling resulting in Hermitian SOS operators described in Eq.~\eqref{eq:sos_op_delta} and Eq.~\eqref{eq:fock_sos_nn_dec_algbra_two_body} is denoted Fock-SOS-nn-dec. For the quadratic part of the Fock-SOS-nn-dec algebra we considered nearest-neighbor orbital sets $(\delta)$ of sizes $[2, 6]$. As the spatial range of the cluster is increased the free-fermion solution at $U=0$ is approached from below. For all cluster sizes a valid SOS representation of the Hamiltonian is constructed. Due to the decoupling of the quadratic and quartic components of the Fock-SOS-nn-dec algebra the lower bound systematically decreases with increasing $U/|t|$ strength. For any spectral gap amplification protocol~\cite{low2025fast, king2025quantum} that relies on the SOS gap being small, this algebra would likely be insufficient. The nearest-neighbor SOS generating algebra corresponding to Fock-SOS-nn (blue curve in Fig.~\ref{fig:fock_space_hub}) follows the exact diagonalization of the model (black curve in Fig.~\ref{fig:fock_space_hub}) at the expense of a substantially more complicated SOS form of the Hamiltonian. The static lower-bound energy beyond $U=6$ is likely due to the nearest-neighbor index constraint on $O_{ij}$.

\begin{figure}[H]
    \centering
    \includegraphics[width=8.5cm]{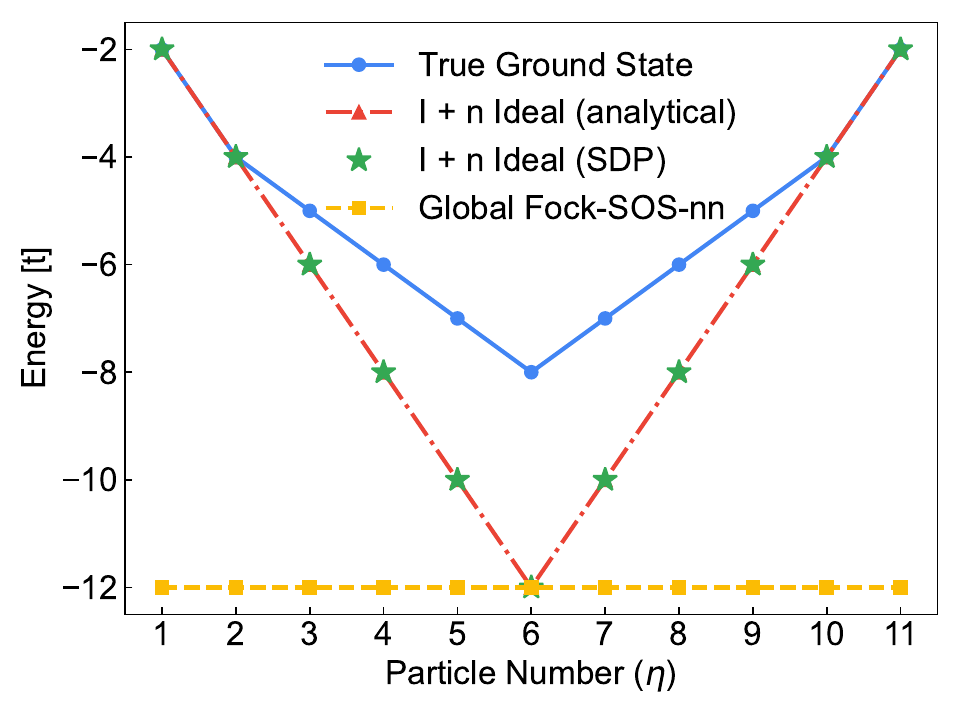}
    \caption{\textcolor{black}{$U=0$ one-dimensional periodic boundary six-site Hubbard model energies versus particle number $\eta$ and lower bounds using the Fock-SOS-nn algebra with the weighted SOS constraint of Eq.~\eqref{eq:weight_in_sos}. The yellow line is the Fock-SOS-nn without the weighting term. In Appendix~\ref{app:hubbard_lower_bounds} we prove this lower bound energy is particle number independent. In red is the Fock-SOS-nn SOS generating algebra with the linear ideal term. We also prove that this lower bound should scale with the particle (or hole) number of the simulation. We plot a numerical verification as green stars which confirm the analytical result.}\label{fig:hilbert_space_hub}}
    \label{fig:lattice}
\end{figure}

In Fig.~\ref{fig:hilbert_space_hub} we plot the $U=0$ SOS solution using the Fock-SOS-nn algebra with the weighted SOS positive ansatz including the particle number polynomial at different particle counts. The blue curve \textcolor{black}{is an exact calculation of the energy} and shows the energy \textcolor{black}{decreases} and then \textcolor{black}{increases} as the filling is changed from 1 electron to 11 electrons. \textcolor{black}{We prove in Appendix~\ref{app:hubbard_lower_bounds} that for the Fock-SOS-nn algebra without an Ideal term the SOS lower bound energy is particle number independent 
\begin{align}
E_{\text{Fock-SOS-nn}} = -2ztL
\end{align}
where $z$ is the lattice dimensional (in this case $z=1$) and $L$ is the total number of lattice sites (in this case $L=6$). We verify this proof numerically and plot the results as the yellow dashed line in Fig.~\ref{fig:hilbert_space_hub}. With the addition of the weighting terms to the SOS we can analytically show (in Appendix~\ref{app:hubbard_lower_bounds}) that we recover a non-constant energy for the lower bound. The analytical derived bounds $E_{\text{Fock-SOS-nn+ideal}}=\max\left(-2tz\eta, \; 2tz\eta - 4Lt\right)$ are shown in red while the numerical verification is shown as a green star. The weighting term in the SOS allows the lower bound to become particle dependent and track the true energy dispersion.}

\subsection{Quartic Hamiltonian}
\label{SEC:QUARTIC_DUAL}
Let us now consider the Coulomb Hamiltonian that arises in non-relativistic electronic structure theory, represented within a basis of spin-orbitals $\{\phi_{i\sigma}\}$:
\begin{align}
\hat{H} = \sum_{ij}\sum_{\sigma} a^\dagger_{i\sigma} a_{j\sigma}  (T_{i\sigma j\sigma} + V_{i\sigma j\sigma}) + \frac{1}{2} \sum_{ijkl} \sum_{\sigma \tau} (ik|jl) a^\dagger_{i\sigma} a^\dagger_{j\tau} a_{l\tau} a_{k\sigma}
\end{align}
or
\begin{align}
\label{EQN:H_CHEMIST}
\hat{H} = \sum_{ij}\sum_{\sigma} a^\dagger_{i\sigma} a_{j\sigma}  h_{i\sigma j\sigma} + \frac{1}{2} \sum_{ijkl} \sum_{\sigma \tau} (ik|jl) a^\dagger_{i\sigma} a_{k\sigma}  a^\dagger_{j\tau} a_{l\tau} 
\end{align}
Here, the labels $i, j, k, l$ refer to the spatial part of the orbital $\phi$, and the labels $\sigma$ and $\tau$ refer to the spin part. The symbols $T_{i\sigma j\sigma}$ and $V_{i\sigma j\sigma}$ refer to electronic kinetic energy integrals and electron-nucleus potential energy integrals, respectively, and $(ik|jl)$ is an electron repulsion integral (ERI)  in chemists' notation. In Eq.~\ref{EQN:H_CHEMIST}, we have rearranged the two-body part of the Hamiltonian and introduced $h_{i\sigma j\sigma} = T_{i\sigma j\sigma} + V_{i\sigma j\sigma} - \frac{1}{2}\sum_{k} (ik|kj)$. For compactness below, we can write this Hamiltonian as a sum of one- and two-body operators: $\hat{H} = \hat{H}^{(1)} + \hat{H}^{(2)}$.

Given the Hamiltonian in Eq.~\ref{EQN:H_CHEMIST}, we can find a lower-bound to the ground-state energy, $E$, by building a sum-of-squares Hamiltonian of the form
\begin{equation}
\label{EQN:H_SOS_QUARTIC}
\hat{H}_{\rm SOS} = \sum_\gamma O^\dagger_\gamma O_\gamma = \hat{H}^{(1)}_\text{SOS} + \hat{H}^{(2)}_\text{SOS} + \lambda 
\end{equation}
Here, $\hat{H}_\text{SOS}^{(1)}$ and $\hat{H}_\text{SOS}^{(2)}$ represent one- and two-body operators, and $\lambda$ is a scalar, the negative of which gives us a lower-bound to the ground-state energy of this Hamiltonian ({\em i.e.}, $-\lambda = E_\text{SOS}$). Since the Hamiltonian is quartic in the creation/annihilation operators, the minimal SOS should be at least a quadratic polynomial in these operators. The full rank-2 SOS is defined by
\begin{align}
O_{\gamma} \in \text{Span}\left(a_{i\sigma}, a_{i\sigma}^{\dagger}, a_{i\sigma}a_{j\tau}, a^\dagger_{i\sigma}a^\dagger_{j\tau}, a^\dagger_{i\sigma} a_{j\tau}, a_{i\sigma} a^\dagger_{j\tau} \right )
\end{align}
As in the lattice Hamiltonian case, if we are not careful, the SOS generated from these operators in Eq.~\ref{EQN:H_SOS_QUARTIC} will include spurious non-particle-conserving and non-spin-conserving terms. We could explicitly force such terms to be zero with linear constraints, but it would be more computationally efficient to use an SOS that decouples the different particle-number and spin sectors. As such, we define the rank-2 SOS Hamiltonian as
\begin{align}
\label{EQN:H_SOS_QUARTIC_BLOCKED}
    \hat{H}_\text{SOS} & = \sum_\gamma O_{G,\gamma}^\dagger O_{G,\gamma} +  \sum_\gamma \sum_{\sigma \neq \tau} O_{G_{\sigma\tau},\gamma}^\dagger O_{G_{\sigma\tau},\gamma} \nonumber \\
    &+ \sum_\sigma \left ( \sum_\gamma O_{D_{\sigma\sigma},\gamma}^\dagger O_{D_{\sigma\sigma},\gamma} + \sum_\gamma O_{Q_{\sigma\sigma},\gamma}^\dagger O_{Q_{\sigma\sigma},\gamma}\right ) \nonumber \\
    &+  \sum_\gamma O_{D_{\alpha\beta},\gamma}^\dagger O_{D_{\alpha\beta},\gamma} + \sum_\gamma O_{Q_{\alpha\beta},\gamma}^\dagger O_{Q_{\alpha\beta},\gamma}
\end{align}
with
\begin{align}\label{EQN:BASIS}
O_{G,\gamma} &= \sum_{\sigma} \sum_{ij}\left ( g^\gamma_{i\sigma j\sigma}a_{j\sigma}^\dagger a_{i\sigma} + \bar{g}^\gamma_{i\sigma j\sigma}a_{j\sigma}a^\dagger_{i\sigma} \right )\\
O_{G_{\sigma\tau},\gamma} &= \sum_{ij} \left ( g^\gamma_{i \sigma j \tau}a_{j \tau}^\dagger a_{i \sigma} + \bar{g}^\gamma_{i \tau j \sigma}a_{j \sigma}a^\dagger_{i \tau} \right );~~~\sigma \neq \tau \\
O_{D_{\sigma\sigma},\gamma} &= \sum_{i < j} d^\gamma_{i\sigma j\sigma} (a_{j\sigma} a_{i\sigma} - a_{i\sigma} a_{j\sigma} ) \\
O_{Q_{\sigma\sigma},\gamma} &= \sum_{i < j} q^\gamma_{i\sigma j\sigma} (a^\dagger_{j\sigma} a^\dagger_{i\sigma} - a^\dagger_{i\sigma} a^\dagger_{j\sigma} ) \\
O_{D_{\alpha\beta},\gamma} &= \sum_{ij} \left (  d^\gamma_{i\alpha j\beta} a_{j\beta} a_{i\alpha} + d^\gamma_{j\beta i\alpha} a_{i\alpha}a_{j\beta} \right );~~~\sigma \neq \tau \\
O_{Q_{\alpha\beta},\gamma} &= \sum_{ij} \left (  q^\gamma_{i\alpha j\beta} a^\dagger_{j\beta} a^\dagger_{i\alpha} + q^\gamma_{j\beta i\alpha}a^\dagger_{i\alpha}a^\dagger_{j\beta} \right );~~~\sigma \neq \tau
\end{align}
Each SOS generator can be used to form a metric matrix.
Detailed mapping constructions for the elements of each metric matrix and how they relate to the Hamiltonian are provided in Appendix~\ref{app:quartic_mapping_conditions}.

In terms of block structure, the coefficient matrix appearing in $\sum_\gamma O_{G_{\gamma}}^\dagger O_{G_{\gamma}}$ has 16 blocks, arranged as
\begin{align}\label{eq:G_same_spin}
{\bf G} = 
    \begin{pmatrix}
{\bf G}^{\alpha\alpha}_{\alpha \alpha} & {\bf G}^{\beta\beta}_{\alpha \alpha} & {\bf G^\prime}^{\alpha\alpha}_{\alpha \alpha} & {\bf G^\prime}^{\beta\beta}_{\alpha \alpha} \\
{\bf G}^{\alpha\alpha}_{\beta\beta} & {\bf G}^{\beta\beta}_{\beta\beta} & {\bf G^\prime}^{\alpha\alpha}_{\beta\beta} & {\bf G^\prime}^{\beta\beta}_{\beta\beta} \\
{\bf G^{\prime\prime}}^{\alpha\alpha}_{\alpha \alpha} & {\bf G^{\prime\prime}}^{\beta\beta}_{\alpha \alpha} & {\bf G^{\prime\prime\prime}}^{\alpha\alpha}_{\alpha \alpha} & {\bf G^{\prime\prime\prime}}^{\beta\beta}_{\alpha \alpha} \\
{\bf G^{\prime\prime}}^{\alpha\alpha}_{\beta\beta} & {\bf G^{\prime\prime}}^{\beta\beta}_{\beta\beta} & {\bf G^{\prime\prime\prime}}^{\alpha\alpha}_{\beta\beta} & {\bf G^{\prime\prime\prime}}^{\beta\beta}_{\beta\beta} \\
    \end{pmatrix} \succeq 0
\end{align}
where the prime notation and spin labels refer to various terms appearing in Eq.~\ref{eqn:Gsos_appendix}. The coefficient matrices appearing in Appendix~\ref{app:quartic_mapping_conditions} and corresponding to $\sum_\gamma O_{G_{\alpha\beta},\gamma}^\dagger O_{G_{\alpha\beta},\gamma}$, $\sum_\gamma O_{G_{\beta\alpha},\gamma}^\dagger O_{G_{\beta\alpha},\gamma}$, $\sum_\gamma O_{D_{\alpha\beta},\gamma}^\dagger O_{D_{\alpha\beta},\gamma}$, and $\sum_\gamma O_{Q_{\alpha\beta},\gamma}^\dagger O_{Q_{\alpha\beta},\gamma}$ have four blocks, arranged as 
\begin{align}
\label{eqn:Gab_sos}
{\bf G}_{\alpha \beta} = 
    \begin{pmatrix}
{\bf G}^{\alpha\beta}_{\alpha \beta} & {\bf G^\prime}^{\beta\alpha}_{\alpha\beta} \\
{\bf G^{\prime\prime}}^{\alpha\beta}_{\beta\alpha} & {\bf G^{\prime\prime\prime}}^{\beta\alpha}_{\beta \alpha} \\
    \end{pmatrix}\succeq 0
\end{align}
\begin{align}
\label{eqn:Gba_sos}
{\bf G}_{\beta \alpha} = 
    \begin{pmatrix}
{\bf G}^{\beta\alpha}_{\beta \alpha} & {\bf G^\prime}^{\alpha\beta}_{\beta\alpha} \\
{\bf G^{\prime\prime}}^{\beta\alpha}_{\alpha\beta} & {\bf G^{\prime\prime\prime}}^{\alpha\beta}_{\alpha \beta} \\
    \end{pmatrix}\succeq 0
\end{align}
\begin{align}
{\bf D}_{\alpha \beta} = 
    \begin{pmatrix}
{\bf D}^{\alpha\beta}_{\alpha \beta} & {\bf D}^{\beta\alpha}_{\alpha\beta} \\
{\bf D}^{\alpha\beta}_{\beta\alpha} & {\bf D}^{\beta\alpha}_{\beta \alpha} \\
    \end{pmatrix}\succeq 0
\end{align}
\begin{align}
\label{eq:Q_opposite_spin}
{\bf Q}_{\alpha \beta} = 
    \begin{pmatrix}
{\bf Q}^{\alpha\beta}_{\alpha \beta} & {\bf Q}^{\beta\alpha}_{\alpha\beta} \\
{\bf Q}^{\alpha\beta}_{\beta\alpha} & {\bf Q}^{\beta\alpha}_{\beta \alpha} \\
    \end{pmatrix}\succeq 0
\end{align}
while the coefficient matrices arising in the remaining operators ($\sum_\gamma O_{D_{\alpha\alpha},\gamma}^\dagger O_{D_{\alpha\alpha},\gamma}$, etc. ) all consist of a single positive semidefinite block. Again, the prime notation and spin labels refer to various terms defined in Appendix~\ref{app:quartic_mapping_conditions}.

The way in which the the elements of ${\bf G}$, ${\bf G}_{\alpha\beta}$, ${\bf G}_{\beta\alpha}$, etc. contribute to the different components of $\hat{H}_\text{SOS}$ is not unique. We can see that this is the case by noting that there are four unique and equally valid ways in which the elements of the ${\bf G}_{\alpha\alpha}^{\alpha\alpha}$ subblock of ${\bf G}$ can map to the two-electron part of the Hamiltonian, {\em i.e.} 
\begin{align}
\label{eqn:indistinguishability_1}
\hat{H}_\text{SOS} &\leftarrow \sum_{ijkl} G^{i\alpha k\alpha}_{l\alpha j\alpha} a^\dagger_{i\alpha} a_{k\alpha} a^\dagger_{j\alpha} a_{l\alpha}  \\
\label{eqn:indistinguishability_2}
\hat{H}_\text{SOS} & \leftarrow \sum_{ijkl} G^{j\alpha k_\alpha}_{l\alpha i\alpha} \left ( a^\dagger_{j\alpha} a_{l\alpha} \delta_{i\alpha, k\alpha} + a^\dagger_{i\alpha} a_{l\alpha} \delta_{k\alpha, j\alpha} - a^\dagger_{i\alpha} a_{k\alpha} a^\dagger_{j\alpha} a_{l\alpha} \right )\\
\label{eqn:indistinguishability_3}
\hat{H}_\text{SOS} & \leftarrow \sum_{ijkl} G^{i\alpha l\alpha}_{k\alpha j\alpha} \left ( a^\dagger_{i\alpha} a_{k\alpha} \delta_{j\alpha, l\alpha} + a^\dagger_{i\alpha} a_{l\alpha} \delta_{k\alpha, j\alpha} - a^\dagger_{i\alpha} a_{k\alpha} a^\dagger_{j\alpha} a_{l\alpha} \right ) \\
\label{eqn:indistinguishability_4}
\hat{H}_\text{SOS} & \leftarrow \sum_{ijkl} G^{j\alpha l\alpha}_{k\alpha i\alpha} \left ( a^\dagger_{j\alpha} a_{k\alpha} \delta_{i\alpha, l\alpha} - a^\dagger_{i\alpha} a_{l\alpha} \delta_{j\alpha, k\alpha} + a^\dagger_{i\alpha} a_{k\alpha} a^\dagger_{j\alpha} a_{l\alpha}  \right ) 
\end{align}
In particular, four elements of ${\bf G}_{\alpha\alpha}^{\alpha\alpha}$ can contribute to the same piece of $\hat{H}_\text{SOS}^{(2)}$. A full accounting of the indistinguishability of electrons requires that we include each of these contributions, as well as all other unique mappings between the SOS and the Hamiltonian that involve the other (sub)blocks of the other matrices in the SOS. The full set of mappings to the Hamiltonian is provided in Appendix~\ref{app:quartic_mapping_conditions}.

It is important to recognize that, by considering all possible ways of grouping contributions from the SOS into $\hat{H}_\text{SOS}^{(1)}$ and $\hat{H}_\text{SOS}^{(2)}$, the SOS Hamiltonian has been fully antisymmetrized. As a consequence, the only way we can ensure that $\hat{H}_\text{SOS}^{(2)} = \hat{H}^{(2)}$ in our program is if the same-spin two-electron contributions to $\hat{H}$ are antisymmetrized as well. In other words, rather than Eq.~\ref{EQN:H_CHEMIST}, we use
\begin{align}
\label{EQN:H_ANTISYM}
\hat{H} &= \sum_{ij}\sum_{\sigma} a^\dagger_{i\sigma} a_{j\sigma}  h_{i\sigma j\sigma} 
+\frac{1}{2} \sum_{ijkl} \sum_{\sigma \neq \tau} (ik|jl) a^\dagger_{i\sigma} a_{k\sigma}  a^\dagger_{j\tau} a_{l\tau} \nonumber \\
&+ \frac{1}{4} \sum_{ijkl} \sum_{\sigma} [(ik|jl)-(il|jk)] a^\dagger_{i\sigma} a_{k\sigma}  a^\dagger_{j\sigma} a_{l\sigma}
\end{align}
with $h_{i\sigma j\sigma} = T_{i\sigma j\sigma} + V_{i\sigma j\sigma} - \frac{1}{4}\sum_k [(ik|kj)-(ij|kk)]$. 
Now, as discussed earlier, the lower bound to the ground-state energy of the quartic Hamiltonian can be obtained from the semidefinite program
\begin{align}
\max~&~-\lambda \nonumber \\
\text{such that}~&~\hat{H}^{(n)}_\text{SOS} = \hat{H}^{(n)} \nonumber \\
\text{and}~&~{\bf G}\succeq 0,~{\bf G}_{\alpha\beta}\succeq 0,~\text{etc.}
\end{align}
\textcolor{black}{Here, $\hat{H}^{(n)}_\text{SOS}$ and $\hat{H}^{(n)}$ refer to the one-body ($n=1)$ and two-body ($n=2$) parts of $\hat{H}_\text{SOS}$ and $\hat{H}$.}

Following the weighted sum-of-squares construction we can now constrain other symmetries by the addition of polynomials of the form given in Eq.~\eqref{eq:wsos_nrep}. In the spinful representation of quadratic order we can set the polynomials $f_{i}$ to any monomial of degree$-1$ or degree$-2$. This involves augmenting the SDP with vectors of variables corresponding to 
\begin{align}
\sum_{i}f_{i}r_{i} + r_{i}f_{i}^{\dagger}
\end{align}
for $\text{deg}(f_{i}) \leq d$. For quadratic generators, this means $f_{i}$ is in the span of particle-conserving quadratic generators
\begin{align}
f_{i} \in \text{Span}\left(\mathbb{1}, a_{i\sigma}^{\dagger}a_{i\sigma}, a_{i\sigma}a_{j\sigma}^{\dagger}\right)
\end{align}
due to the Hamiltonian being particle and spin conserving. We also restrict $f_{i}$ to real-valued polynomials. The $f_{i}$ polynomials correspond to operators of the first line in Eq.~\eqref{EQN:BASIS} meaning it touches every block in Eq.~\eqref{eq:G_same_spin}. For example, one polynomial could be
\begin{align}
f_{i}(c, c_{ij,\sigma}, \overline{c}_{ij,\sigma})r_{i} + r_{i}f_{i}^{\dagger}(c, c_{ij\sigma}, \overline{c}_{ij\sigma}) =& \left(c\mathbb{1} + \sum_{\sigma}\sum_{ij}c_{ij, \sigma}a_{i\sigma}^{\dagger}a_{j\sigma} + \sum_{\sigma}\sum_{ij}\overline{c}_{ij, \sigma}a_{i\sigma}a_{j\sigma}^{\dagger}\right) \left( \sum_{i\sigma}a_{i\sigma}^{\dagger}a_{i\sigma}  - \eta \mathbb{1}\right) + \text{h.c.} \\
=& c\sum_{i\sigma}a_{i\sigma}^{\dagger}a_{i\sigma} - c\eta \mathbb{1}  + \sum_{ijk\sigma\tau}c_{ij\sigma}a_{i\sigma}^{\dagger}a_{j\sigma}a_{k\tau}^{\dagger}a_{k\tau}  -\eta\sum_{ij\sigma}c_{ij\sigma}a_{i\sigma}^{\dagger}a_{j\sigma} \nonumber \\
+&  \sum_{ijk\sigma\tau}\overline{c}_{ij\sigma}a_{i\sigma}a_{j\sigma}^{\dagger}a_{k\tau}^\dagger a_{k\tau} - \eta \sum_{ij\sigma}\overline{c}_{ij\sigma}a_{i\sigma}a_{j\sigma}^{\dagger} + \text{h.c.}
\end{align} 
If we do not wish to retain the identity in the algebra, we can replace the it with the anticommutation relation.
\begin{align}
\mathbb{1} = a_{i\sigma}a_{i\sigma}^{\dagger} + a_{i\sigma}^{\dagger}a_{i\sigma}
\end{align}
In total, we add $8 N^{2} + 2$ non-negative variables to the SDP in the form of $1\times 1$ blocks where $N$ is the size of the spatial one-particle basis. 

\subsection{Spin-Free Formalism}
\label{SEC:SPIN_FREE_DUAL}

Here we define the spin-free (SF) formalism of the dual SOS to reduce the number of variables at the expense of the tightness of the lower-bound.
Consider the following SF particle-hole and hole-particle generators 
\begin{align}
\label{EQN:SPIN_FREE_G2_BASIS}
O_{G_\text{SF},\gamma} &=  \sum_{ij} \left ( g^\gamma_{ij} \sum_{\sigma} a_{j\sigma}^\dagger a_{i\sigma}  + \bar{g}^\gamma_{ij} \sum_{\sigma} a_{j\sigma}a^\dagger_{i\sigma} \right ).
\end{align}
As the resulting SOS lacks sufficient flexibility to describe the electronic Hamiltonian, we augment it with the rank-1 generators Eqs.~\ref{eq:O_ladder_D} and~\ref{eq:O_ladder_Q} introduced earlier 
\begin{align}
\label{EQN:D1_Q1}
O_{D_\sigma,\gamma} &=  \sum_{i}  d^\gamma_{i\sigma} a_{i\sigma}  \\
O_{Q_\sigma,\gamma} &=  \sum_{i}  q^\gamma_{i\sigma} a^\dagger_{i\sigma} 
\end{align}
and the SOS Hamiltonian is
\begin{align}
\label{EQN:HSOS}
    \hat{H}_\text{SOS} & = \sum_\gamma O_{G_\text{SF},\gamma}^\dagger O_{G_\text{SF},\gamma} + \sum_\sigma \left ( \sum_\gamma O_{D_{\sigma},\gamma}^\dagger O_{D_{\sigma},\gamma} + \sum_\gamma O_{Q_{\sigma},\gamma}^\dagger O_{Q_{\sigma},\gamma}\right )
\end{align}
with
\begin{align}
\label{EQN:SPIN_FREE_OPERATORS}
    \sum_\gamma O_{G_\text{SF},\gamma}^\dagger O_{G_\text{SF},\gamma}  &=  \sum_{ijkl} \left ( G^{ik}_{lj}  \sum_{\sigma\tau} \right . a^\dagger_{i\sigma}a_{k\sigma}a^\dagger_{j\tau}a_{l\tau} + {G^{\prime}}^{ik}_{jl}  \sum_{\sigma\tau} a^\dagger_{i\sigma}a_{k\sigma}a_{l\tau} a^\dagger_{j\tau} \nonumber \\
    &+ {G^{\prime\prime}}^{ki}_{lj}  \sum_{\sigma\tau} a_{k\sigma}a^\dagger_{i\sigma} a^\dagger_{j\tau}a_{l\tau} + {G^{\prime\prime\prime}}^{ki}_{jl}  \left . \sum_{\sigma\tau} a_{k\sigma}a^\dagger_{i\sigma}a_{l\tau} a^\dagger_{j\tau} \right ) \\
    \sum_\gamma O_{D_\sigma,\gamma}^\dagger O_{D_\sigma,\gamma} &= \sum_{ij} D^{i\sigma}_{j\sigma} a^\dagger_{i\sigma} a_{j\sigma} \\
    \sum_\gamma O_{Q_\sigma,\gamma}^\dagger O_{Q_\sigma,\gamma} &= \sum_{ij} Q^{j\sigma}_{i\sigma} a_{j\sigma} a^\dagger_{i\sigma}.
\end{align}
The coefficient matrix appearing in $\sum_\gamma O_{G_\text{SF},\gamma}^\dagger O_{G_\text{SF},\gamma}$ has four blocks, arranged as
\begin{align}
{\bf G_\text{SF}} = 
    \begin{pmatrix}
{\bf G} & {\bf G^\prime} \\
{\bf G^{\prime\prime}} & {\bf G^{\prime\prime\prime}} \\
    \end{pmatrix} \succeq 0
\end{align}
The remaining coefficient matrices consist of a single positive semidefinite block. 
After bringing the creation and annihilation operators in Eqs.~\ref{EQN:SPIN_FREE_OPERATORS} to a common order, we can see that 
\begin{align}
\hat{H}^{(2)}_\text{SOS} &= \sum_{ijkl} (G^{ik}_{lj} - {G^{\prime}}^{ik}_{jl} - {G^{\prime\prime}}^{ki}_{lj} + {G^{\prime\prime\prime}}^{ki}_{jl} ) \sum_{\sigma\tau} a^\dagger_{i\sigma} a_{k\sigma} a^\dagger_{j\tau} a_{l\tau} \\
\hat{H}^{(1)}_\text{SOS} &= 2 \sum_{ij} \sum_{\sigma} \left ( \sum_p \left ( {G^{\prime}}^{ij}_{pp} + {G^{\prime\prime}}^{pp}_{ji} - {G^{\prime\prime\prime}}^{pp}_{ij} - {G^{\prime\prime\prime}}^{ji}_{pp} \right ) + \frac{1}{2} \left ( D^{i\sigma}_{j\sigma} - Q^{j\sigma}_{i\sigma} \right )   \right )  a^\dagger_{i\sigma} a_{j\sigma} \\
\lambda &= 4 \sum_{pq} {G^{\prime\prime\prime}}^{pp}_{qq} + \sum_p \sum_\sigma Q^{p\sigma}_{p\sigma}
\end{align}
Note that two-electron terms for a given $i$, $j$, $k$, and $l$ have the same coefficient regardless of the spins associated with the labels. As a result, this spin-free formalism cannot be used with the antisymmetrized Hamiltonian in Eq.~\ref{EQN:H_ANTISYM}, which uses antisymmetrized versus bare two-electron integrals in the same-spin and opposite spin parts, respectively. Rather, the spin-free formalism must be applied to the Hamiltonian in Eq.~\ref{EQN:H_CHEMIST}.

We can further reduce the number of variables without impacting the tightness of the lower bound by introducing spin-free unitary group generator $E_{ij}$ and hole rotation generators $\overline{E}_{kl}$
\begin{align}
E_{ij} = \sum_{\sigma}a_{i\sigma}^{\dagger}a_{j\sigma},\quad \overline{E}_{kl}=\sum_{\tau}a_{k\tau}a_{l\tau}^{\dagger}
\end{align}
which follow the commutation relations
\begin{align}
\left[E_{ij}, E_{kl} \right] = \delta_{jk}E_{il} - \delta_{il}E_{kj} \\
\left[\overline{E}_{ij}, \overline{E}_{kl} \right] = \delta_{jk}\overline{E}_{il} - \delta_{il}\overline{E}_{kj} \\
\left[E_{ij}, \overline{E}_{kl} \right] = 2\delta_{ik}\delta_{jl} + \delta_{ik}E_{lj} + \delta_{jl}\overline{E}_{ki} \\
E_{ij} = 2\delta_{ij} - \overline{E}_{ji} \label{eq:delta_resolve_1}
\end{align}
The relationship between $E_{ij}$ and $\bar{E}_{ij}$ in the last expression suggests that the latter generators can be excluded from the algebra via the introduction of the unit operator, {\em i.e.}, 
\begin{align}
O_{G_\text{SF},\gamma} &=  g^\gamma_1 \hat{1} + \sum_{ij} g^\gamma_{ij} \sum_{\sigma} a_{j\sigma}^\dagger a_{i\sigma}  
\end{align}
Now, we have
\begin{align}
\hat{H}^{(2)}_\text{SOS} &= \sum_{ijkl} G^{ik}_{lj} \sum_{\sigma\tau} a^\dagger_{i\sigma} a_{k\sigma} a^\dagger_{j\tau} a_{l\tau} \label{eq:spinfree_efficient_H2}\\
\hat{H}^{(1)}_\text{SOS} &= \sum_{ij} \sum_{\sigma} \left ( {G}^{1}_{ji} + {G}^{ij}_{1} + D^{i\sigma}_{j\sigma} - Q^{j\sigma}_{i\sigma} \right )   a^\dagger_{i\sigma} a_{j\sigma} \label{eq:spinfree_efficient_H1} \\
\lambda &= {G}^{1}_{1} + \sum_p \sum_\sigma Q^{p\sigma}_{p\sigma} \label{eq:spinfree_efficient_I}
\end{align}
where the {\bf G} matrix still has four blocks, arranged as
\begin{align}\label{eq:gsf}
{\bf G_\text{SF}} = 
    \begin{pmatrix}
{\bf G} & {\bf G}^1 \\
{\bf G}_1 & G^1_1 \\
    \end{pmatrix} \succeq 0
\end{align}
As with the larger spin-free algebra, the {\bf G} coefficient matrix in this form contributions to both the one- and two-electron parts of the Hamiltonian, as well as the lower bound. The larger spin-free algebra produces an identical lower bound to the algebra smaller algebra that includes an identity operator because any element of the larger algebra SOS generator can be re-expressed as an element of the smaller algebra. This result was numerically verified in previous work~\cite{low2025fast}.

Finally, we can augment this spin-free algebra SOS with linear constraints to utilize a weighted sum-of-squares set.  Due to the structure of the algebra the free variables now have the form
\begin{align}
f_{i} \in \text{Span}\left(\mathbb{1}, E_{ij}\right)
\end{align}
where 
\begin{align}
f_{i}^{\dagger}r_{j} + r_{j}^{\dagger}f_{i} =& \left(c \mathbb{1} + \sum_{ij}c_{ij}E_{ij} \right)\left( \sum_{i}E_{ii} - \eta\mathbb{1}\right) + \left( \sum_{i}E_{ii} - \eta\mathbb{1}\right) \left(c \mathbb{1} + \sum_{ij}c_{ij}E_{ji} \right) \\
=&  c\sum_{i}E_{ii}  - c\eta\mathbb{1} + \sum_{ijk}c_{ij}E_{ij}E_{kk} -\eta \sum_{ij}c_{ij}E_{ij} \nonumber \\
 &+  c\sum_{i}E_{ii} + \sum_{kij}c_{ij}E_{kk}E_{ji} - c\eta\mathbb{1} - \eta\sum_{ij}c_{ij}E_{ji} \\
 =& 2c\sum_{i}E_{ii} - 2c\eta\mathbb{1} + \sum_{ijk}c_{ij}\left(E_{ij}E_{kk} + E_{kk}E_{ji}\right) - 2\eta\sum_{ij}c_{ij}\left(E_{ij} + E_{ji}\right)
\end{align}
which adds $2N^{2} + 2$ more non-negative variables to the SDP as $1\times1$ blocks. 
We augment the expression for $\lambda$ in Eq.~\eqref{eq:spinfree_efficient_I} with 
\begin{align}
\lambda \mathrel{+}= -2c\eta 
\end{align}
and augment the linear constraints in Eq.~\eqref{eq:spinfree_efficient_H1} and Eq.~\eqref{eq:spinfree_efficient_H2} with the following expressions:
\begin{align}
H_{\text{SOS}}^{(1)} \mathrel{+}=& 2c\sum_{i\sigma}a_{i\sigma}^{\dagger}a_{i\sigma} - 2\eta\sum_{ij\sigma}c_{ij}\left(a_{i\sigma}^{\dagger}a_{j\sigma} + a_{j\sigma}^{\dagger}a_{i\sigma}\right) \\
H_{\text{SOS}}^{(2)} \mathrel{+}=&  \sum_{ijk}c_{ij}\left(E_{ij}E_{kk} + E_{kk}E_{ji}\right)
\end{align}

\section{The v2RDM primal program relationship to the SOS dual}
We now consider the v2RDM primal problem, which is closely related to the dual problem outlined in the previous section. We begin by setting up a simpler problem where we minimize the energy of a system of non-interacting electrons with respect to variations in the 1RDM, subject to 1-particle $n$-representability conditions. This system is describable by a quadratic Hamiltonian of the form
\begin{align}
 \label{EQN:H_QUADRATIC}
 \hat{H} = \sum_{ij} h_{ij} \sum_\sigma  a^\dagger_{i\sigma} a_{j\sigma}
\end{align}
where the coefficient matrix, $h_{ij}$, encodes the kinetic energy of and potential experienced by the particles.
The ground-state energy for the system is obtained from the primal problem
\begin{align}
\label{EQN:PRIMAL_QUADRATIC}
&\text{min}~\text{Tr}(\bf H x ) \nonumber \\
&\text{such that}~\text{Tr}({\bf A}_I{\bf x}) - b_I = 0~ \forall~ \textcolor{black}{\text{constraints, }}I \nonumber \\
&\text{and}~{\bf x} \succeq 0
\end{align}
Here, the primal solution, ${\bf x}$ is
\begin{align}
{\bf x} = \begin{pmatrix}
x_{1} & 0 & 0 & 0 & 0\\
0 & {}^1{\bf D}_\alpha & 0 & 0 & 0 \\
0 & 0 & {}^1{\bf D}_\beta & 0 & 0 \\
0 & 0 & 0 & {}^1{\bf Q}_\alpha & 0 \\
0 & 0 & 0 & 0 & {}^1{\bf Q}_\beta
\end{pmatrix}
\end{align}
where ${}^1{\bf D}_\sigma$ and ${}^1{\bf Q}_\sigma$ represent the 1RDM and the one-hole RDM, with elements
\begin{align}
{}^1D^{i\sigma}_{j\sigma} = \langle \Psi | a^\dagger_{i\sigma} a_{j\sigma}| \Psi \rangle \\
{}^1Q^{i\sigma}_{j\sigma} = \langle \Psi | a_{i\sigma} a^\dagger_{j\sigma}| \Psi \rangle 
\end{align}
and $x_1$ is a scalar term that is constrained to be equal to one.
The Hamiltonian matrix is arranged as
\begin{align}
{\bf H} = \begin{pmatrix}
0 & 0 & 0 & 0 & 0 \\
0 & {\bf h} & 0 & 0 & 0 \\
0 & 0 & {\bf h} & 0 & 0 \\
0 & 0 & 0 & 0 & 0 \\
0 & 0 & 0 & 0 & 0 
\end{pmatrix}
\end{align}
where the elements of ${\bf h}$ are $h_{ij}$ from Eq.~\ref{EQN:H_QUADRATIC}. The matrices ${\bf A}_I$ and scalars $b_I$ encode the $n$-representability conditions for the problem\textcolor{black}{, with $I$ representing a given constraint on the RDMs}. The primal program for the quadratic Hamiltonian is exact because the complete ensemble $n$-representability conditions for the 1RDM are known.  We have three types of conditions: 1) the scalar term is constrained as
\begin{align}
\label{EQN:ONE_EQUALS_ONE}
x_1 = 1
\end{align}
2) there are two conditions that specify the particle number and $z$-projection of spin
\begin{align}
\label{EQN:TRACE_D1_EQUALS_N}
\text{Tr}({}^1{\bf D}_\sigma) - n_\sigma x_1 = 0
\end{align}
where $\sigma \in \{\alpha, \beta\}$ and $n_\sigma$ is the number of $\sigma$-spin electrons, 3) there are two sets conditions that must be satisfied by the algebra of fermionic operators, {\em i.e.}
\begin{align}
\label{EQN:D1_PLUS_Q1_EQUALS_DELTA}
{}^1D^{i\sigma}_{j\sigma} + {}^1Q^{j\sigma}_{i\sigma} - \delta_{ij} x_1 = 0
\end{align}
where, again, $\sigma \in \{\alpha, \beta\}$. We include the scalar in these expressions so that the problem can be represented in standard form, {\em i.e.}, where the first constraint evaluates to one ($b_0 = 1$), and the remaining constraints evaluate to zero ($b_{I>0} = 0$). 

In this standard form, the Lagrangian for the primal problem is
\begin{align}
\mathcal{L}({\bf x}, y, {\bf B}) = \text{Tr}\left ( \left [ {\bf H} - \sum_{I} y_I {\bf A}_I - {\bf B}\right ] {\bf x} \right ) + y_0
\end{align}
where $y_I$ is the Lagrange multiplier for the $I$th linear constraint, and ${\bf B}$ is the Lagrange multiplier for the positive semidefinite constraint on ${\bf x}$,
\begin{align}
{\bf B} = \begin{pmatrix}
B(1) & 0 & 0 & 0 & 0\\
0 & {\bf B}({}^1{D}_\alpha) & 0 & 0 & 0 \\
0 & 0 & {\bf B}({}^1{D}_\beta) & 0 & 0 \\
0 & 0 & 0 & {\bf B}({}^1{Q}_\alpha) & 0 \\
0 & 0 & 0 & 0 & {\bf B}({}^1{Q}_\beta)
\end{pmatrix} \succeq 0
\end{align}
We next define the dual function, which minimizes the Lagrangian with respect to the primal variable
\begin{align}
\mathcal{F}(y, {\bf B}) = \min_{\bf x} \text{Tr}\left ( \left [ {\bf H} - \sum_{I} y_I {\bf A}_I - {\bf B}\right ] {\bf x} \right ) + y_0
\end{align}
which is bounded from below ({\em i.e.}, $\mathcal{F}(y, {\bf B}) \to -\infty$) unless 
\begin{align}
\label{EQN:DUAL_CONSTRAINT}
{\bf H} = {\bf B} + \sum_{I} y_I {\bf A}_I  
\end{align}
Equation \ref{EQN:DUAL_CONSTRAINT} is the dual constraint for the primal problem. As each linear constraint ${\bf A}_{I}$ corresponds to an operator algebra relation, Eq.~\eqref{EQN:DUAL_CONSTRAINT} is exactly the weighted SOS form one would start from in the dual SOS picture assuming that the Lagrange multiplier algebra is $\{\mathbb{1}\}$.

To illustrate the connection more clearly, we demonstrate that under certain scenarios it is possible to convert a weighted SOS into a global SOS. In other words, 
we show that it is possible to extract an SOS Hamiltonian (contained with $\mathcal{B}_{\text{SOS}}$) from the dual constraint (which is in $\mathcal{B}_{\text{wSOS}}$), {\em i.e.}, ${\bf H}_\text{SOS} = {\bf H} - E_\text{primal}$, where $E_\text{primal} = \text{Tr}(\bf{H}\bf{x})$. The operator form of ${\bf H}$ is
\begin{align}
\label{EQN:SOS_OPERATOR_FROM_DUAL}
\hat{H} = \hat{B} + \sum_{I} y_I \hat{A}_I
\end{align}
where $\hat{B}$ and $\hat{A}_i$ are the operator representations of ${\bf B}$ and ${\bf A}_I$. The operator $\hat{B}$ is
\begin{align}
\hat{B} = B(1) + \sum_\sigma \sum_{ij} B\left({}^1D_\sigma\right)^i_j a^\dagger_{i\sigma} a_{j\sigma} + \sum_\sigma \sum_{ij} B({}^1Q_\sigma)^{i\sigma}_{j\sigma} a_{i\sigma} a^\dagger_{j\sigma}
\end{align}
and this operator is in SOS form because, by construction, ${\bf B}\succeq 0$.
Now, let us define the operators, $\hat{A}_I$, corresponding to the linear constraints. For the constraint that $x_1 = 1$, we have
\begin{align}
\label{EQN:OPERATOR_ONE_EQUALS_ONE}
\hat{A}_0 = \hat{1}
\end{align}
where we use the unit operator to emphasize that this constraint touches scalar part of the variational space.
For the trace constraints, we have
\begin{align}
\label{EQN:TRACE_CONSTRAINTS}
\hat{A}_{\text{Tr},\sigma} = \sum_i a^\dagger_{i\sigma} a_{i\sigma} - n_\sigma \hat{1}
\end{align}
Next, we have the constraints arising from the fermionic anticommutation relations, {\em i.e.},
\begin{align}
\label{EQN:OPERATOR_D1_Q1}
\hat{A}_{ij,\sigma} = a^\dagger_{i\sigma}a_{j\sigma} + a^\dagger_{j\sigma}a_{i\sigma} - \delta_{ij}\hat{1}
\end{align}
$\hat{H}$ for the quadratic problem is thus
\begin{align}
\hat{H} &= \hat{B} + \sum_\sigma y_{\text{Tr},\sigma} \hat{A}_{\text{Tr},\sigma} + \sum_{\sigma}\sum_{ij} y_{ij,\sigma} \hat{A}_{ij,\sigma} +y_0 \hat{1} \\
& = \hat{B} + \sum_\sigma y_{\text{Tr},\sigma} (\sum_i a^\dagger_{i\sigma} a_{i\sigma} - n_\sigma \hat{1} ) + \sum_{\sigma}\sum_{ij} y_{ij,\sigma} (a^\dagger_{i\sigma} a_{j\sigma} + a_{j\sigma} a^\dagger_{i\sigma} - \hat{1}) + y_0 \hat{1} 
\end{align}
Collecting terms of each operator type, we have
\begin{align}
\label{EQN:SOS_FROM_DUAL_1}
\hat{H} &= B(1) - \sum_\sigma y_{\text{Tr},\sigma} n_\sigma - \sum_\sigma \sum_i y_{ii,\sigma} + y_0 \nonumber \\
&+ \sum_\sigma \sum_{ij} \left [ B({}^1D_\sigma)^{i}_{j} + y_{\text{Tr},\sigma} \delta_{ij} + y_{ij,\sigma} \right ] a^\dagger_{i\sigma} a_{j\sigma} \nonumber \\
&+ \sum_\sigma \sum_{ij} \left [ B({}^1Q_\sigma)^i_j + y_{ij,\sigma} \right ] a_{i\sigma} a^\dagger_{j\sigma} 
\end{align}
If $y_{\text{Tr},\sigma} \ge 0$ and the matrices defined by $y_{ij,\sigma}$ are positive semidefinite, then the coefficient matrices associated with the particle-hole and hole-particle transition operators in Eq.~\ref{EQN:SOS_FROM_DUAL_1} will be positive semidefinite. If the scalar part exceeds $E_\text{primal}$, then $\hat{H}_\text{SOS} = \hat{H} - E_\text{primal}$ constructed in this way will indeed be an SOS in $\mathcal{B}_{\text{SOS}}$. Unfortunately, we have neither a guarantee that the constant part of $\hat{H}$ exceeds $E_\text{primal}$ nor that the coefficient matrices comprising $\sum_I y_i \hat{A}_I$ are positive semidefinite. As such, our only recourse is to devise a numerical procedure that constructs $\hat{H}_\text{SOS}$ from Eq.~\ref{EQN:SOS_FROM_DUAL_1}. 

We rewrite ${\bf H}$ as
\begin{align}
{\bf H} = \begin{pmatrix}
H(1) & 0 & 0 & 0 & 0 \\
0 & {\bf H}({}^1D_\alpha) & 0 & 0 & 0\\
0 & 0 & {\bf H}({}^1D_\beta) & 0 & 0 \\
0 & 0 & 0 & {\bf H}({}^1Q_\alpha) & 0 \\
0 & 0 & 0 & 0 & {\bf H}({}^1Q_\beta) \\
\end{pmatrix} = {\bf H}_+ + {\bf H}_-
\end{align}
where
\begin{align}
H(1) &= B(1) - \sum_\sigma y_{\text{Tr},\sigma} n_\sigma - \sum_\sigma \sum_i y_{ii,\sigma} + y_0 \\
H({}^1D_\sigma)^{i}_{j} &= B({}^1D_\sigma)^{i}_{j} + y_{\text{Tr},\sigma} \delta_{ij} + y_{ij,\sigma} \\
H({}^1Q_\sigma)^{i}_{j} &= B({}^1Q_\sigma)^{i}_{j} + y_{ij,\sigma} 
\end{align}
and ${\bf H}_+$ and ${\bf H}_-$ represent the positive and negative components of ${\bf H}$, respectively, which can be isolated by diagonalizing ${\bf H}$. The negative parts of ${\bf H}$ can be pushed to complementary spaces the fermionic anticommutation relations, in which case, the Hamiltonian can be expressed as
\begin{align}
\hat{H} &= H(1) + \sum_{\sigma}\sum_i \left [ H_-({}^1D_\sigma)^{i}_{i} + H_-({}^1Q_\sigma)^{i}_{i} \right ] \nonumber \\
& + \sum_\sigma \sum_{ij} \left [ H_+({}^1D_\sigma)^i_j - H_-({}^1Q_\sigma)^j_i \right ] a_{i\sigma}^\dagger a_{j\sigma} \nonumber \\
& + \sum_\sigma \sum_{ij} \left [ H_+({}^1Q_\sigma)^i_j - H_-({}^1D_\sigma)^j_i \right ] a_{i\sigma} a^\dagger_{j\sigma}
\end{align}
With this form, we can finally define $\hat{H}_\text{SOS} = \hat{H} - E_\text{primal}$. If the spin and particle number sectors specified by the constraints in Eq.~\ref{EQN:TRACE_CONSTRAINTS} correspond to the ground state of $\hat{H}$, then $\hat{H}_\text{SOS}$ constructed in this way will be equivalent to that obtained from the direct solution of the dual SOS problem. 
For any other spin or particle-number state, though, $\hat{H}_\text{SOS}$ obtained in this way or via the dual approach could differ. 

Now, we return to the Coulomb Hamiltonian  (Eq.~\ref{EQN:H_CHEMIST}) and consider the simplest possible algebra for the SOS dual problem, which is based on the spin-free formulation of the G matrix. We have a similar program as for the quadratic Hamiltonian, {\em i.e.}, 
\begin{align}
&\text{min}~\text{Tr}(\bf H x ) \nonumber \\
&\text{such that}~\text{Tr}({\bf A}_I{\bf x}) - b_I = 0~ \forall~ I \nonumber \\
&\text{and}~{\bf x} \succeq 0
\end{align}
but, in this case, the primal solution is
\begin{align}
\label{EQN:SPIN_FREE_PRIMAL}
{\bf x} = \begin{pmatrix}
{}^2{\bf G}^\prime & 0 & 0 & 0 & 0 \\ 
0 & {}^1{\bf D}_\alpha & 0 & 0 & 0 \\
0 & 0 & {}^1{\bf D}_\beta & 0 & 0 \\
0 & 0 & 0 & {}^1{\bf Q}_\alpha & 0 \\
0 & 0 & 0 & 0 & {}^1{\bf Q}_\beta \\
\end{pmatrix}
\end{align}
where ${}^2{\bf G}^\prime$ is a block matrix of the form
\begin{align}
\label{EQN:SPIN_FREE_G2_PRIMAL}
{}^2{\bf G}^\prime = \begin{pmatrix}
x_1 & \sum_\sigma \text{vec}\left ( {}^1{\bf D}_\sigma \right ) ^T \\
\sum_\sigma \text{vec}\left ( {}^1{\bf D}_\sigma \right ) & {}^2{\bf G} \\
\end{pmatrix}
\end{align}
In Eqs.~\ref{EQN:SPIN_FREE_PRIMAL} and \ref{EQN:SPIN_FREE_G2_PRIMAL}, ${}^1{\bf D}_\sigma$ and $x_1$ are the same spin-blocks of the 1RDM and constant factor that arose in the quadratic problem, and the elements of ${}^2{\bf G}$ are
\begin{align}
{}^2G^{ij}_{kl} = \langle \Psi | E_{ij} E_{kl}^\dagger | \Psi \rangle
\end{align}
The Hamiltonian matrix is arranged as
\begin{align}
{\bf H} = \begin{pmatrix}
{\bf g}^\prime & 0 & 0 & 0 & 0 \\
0 & {\bf h}_\alpha & 0 & 0 & 0 \\
0 & 0 & {\bf h}_\beta & 0 & 0 \\
0 & 0 & 0 & 0 & 0 \\
0 & 0 & 0 & 0 & 0 \\
\end{pmatrix}
\end{align}
where ${\bf h}_\sigma$ is the matrix representation of the one-body part of the chemistry Hamiltonian in Eq.~\ref{EQN:H_CHEMIST}, and 
\begin{align}
{\bf g}^\prime = \begin{pmatrix}
0 & 0 \\
0 & {\bf g}
\end{pmatrix}
\end{align}
with $g^{ij}_{kl} = \frac{1}{2} (ij|kl)$. 

As in the primal problem involving the quadratic Hamiltonian, ${\bf A}_I$ and $b_I$ encode the $n$-representability conditions for the problem. We have the same constraints as in the quadratic case ({\em i.e.}, Eqs.~\ref{EQN:ONE_EQUALS_ONE}--\ref{EQN:D1_PLUS_Q1_EQUALS_DELTA}), as well as additional constraints to ensure the appropriate relationships between ${}^1{\bf D}_\sigma$ and ${}^2{\bf G}^\prime$ in Eq.~\ref{EQN:SPIN_FREE_G2_PRIMAL}. Note that these constraints represent only a subset of two-particle ensemble $n$-representability conditions, so the primal energy, $E_\text{primal} = \text{Tr}({\bf H}{\bf x})$, will be a lower-bound to the true energy for the target state. 

As above, we have a dual constraint representation of the chemistry Hamiltonian in either matrix or operator form (Eq.~\ref{EQN:DUAL_CONSTRAINT} or \ref{EQN:SOS_OPERATOR_FROM_DUAL}, respectively). The matrix ${\bf H}$ defined by Eq.~\ref{EQN:DUAL_CONSTRAINT} has the form
\begin{align}
{\bf H} = \begin{pmatrix}
{\bf H}({}^2{G}^\prime) & 0 & 0 & 0 & 0\\
0 & {\bf H}({}^1{D}_\alpha) & 0 & 0 & 0 \\
0 & 0 & {\bf H}({}^1{D}_\beta) & 0 & 0 \\
0 & 0 & 0 & {\bf H}({}^1{Q}_\alpha) & 0 \\
0 & 0 & 0 & 0 & {\bf H}({}^1{Q}_\beta)
\end{pmatrix} 
\end{align}
where ${\bf H}({}^2G^\prime)$ has the same structure as ${}^2{\bf G}^\prime$ in Eq.~\ref{EQN:SPIN_FREE_G2_PRIMAL}. The matrices ${\bf B}$ and $\sum_I y_I {\bf A}_I$ that define ${\bf H}$ have the same structure. At this point, we could imagine attempting to construct an SOS Hamiltonian operator from ${\bf H} - E_\text{primal}$, playing similar games as in the quadratic case, {\em e.g.}, by shuffling terms between complimentary spaces via the fermionic anticommutation relations. However, we would run into the same problem as earlier because there is no obvious protocol for constructing an SOS Hamiltonian in this way. In the quadratic case, the SOS Hamiltonian could be determined by isolating the positive and negative components of ${\bf H} = {\bf B} + \sum_I y_I {\bf A}_I$ via diagonalization. Here, however, the problem is more complex because of the way that the two-particle space couples to the constant part of the algebra [{\em i.e.}, in the off-diagonal blocks of ${\bf H}({}^2G^\prime)$]. In fact, the set inclusion in Eq.~\eqref{eq:cone_subset} tells us that expressing the original v2RDM problem in $\mathcal{B}_{\text{SOS}}$ is not possible in all cases. 

\section{Numerical demonstration of lower-bound behavior}

In this Section, we demonstrate numerically that the primal (v2RDM) and dual (SOS) problems laid out in the previous sections yield lower bounds to the ground-state energies of many-electron systems. The v2RDM calculations enforce two-particle (DQG) ensemble $n$-representability conditions,\cite{Percus64_1756} and the SOS Hamiltonians are optimized using either a comparable algebra ({\em i.e.}, the full rank-2 algebra in Sec.~\ref{SEC:QUARTIC_DUAL}) or a subset of that algebra. The SOS optimizations do not include any particle-number or spin-symmetry constraints, {\em i.e.}, they correspond to the standard SOS dual formulation, rather than the ``weighted'' SOS. The semidefinite program for optimizing the SOS was implemented in Python using the \texttt{libSDP} library\cite{libsdp, DePrince24_e1702} of semidefinite programming solvers. The one- and two-electron integrals defining the electronic Hamiltonian in the SOS optimization were taken from the PySCF package.\cite{Chan20_024109} All v2RDM calculations were performed using \texttt{hilbert}\cite{hilbert}, which is a plugin to the \textsc{Psi4}\cite{Sherrill20_184108} package. Reference energies representing the true ground states of the Hamiltonian were obtained from full configuration interaction (CI) calculations, which were performed using PySCF. 

\begin{figure*}[htbp]
    \centering
    \includegraphics{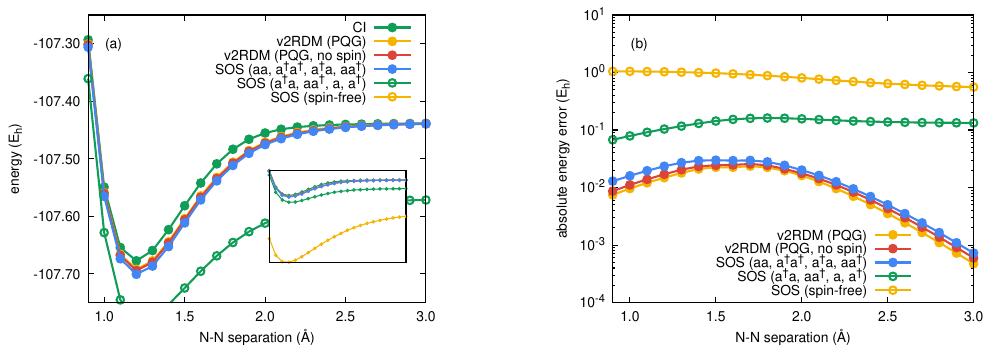}
    \caption{The (a) potential energy curves for the dissociation of molecular nitrogen and (b) errors in the dissociation curves relative to full CI.}
    \label{FIG:N2}
\end{figure*}

\begin{figure*}[htbp]
    \centering
    \includegraphics{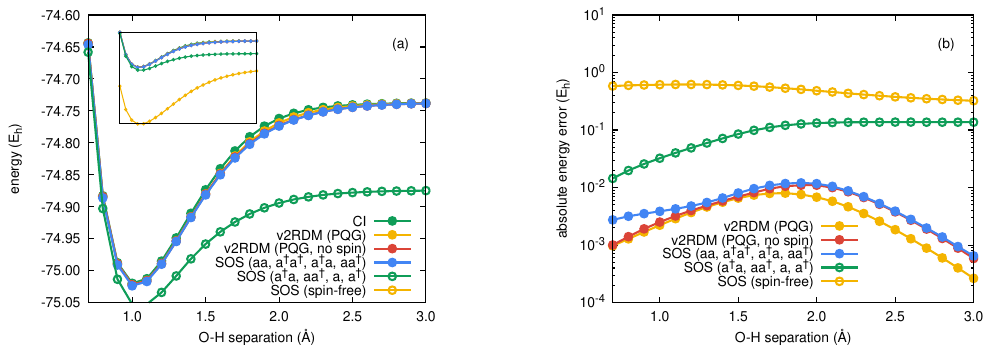}
    \caption{The (a) potential energy curves for the symmetric double dissociation of water and (b) errors in the dissociation curves relative to full CI.}
    \label{FIG:H2O}
\end{figure*}

Figures \ref{FIG:N2} and \ref{FIG:H2O} illustrate potential energy curves (PECs) for the dissociation of molecular nitrogen and the symmetric double dissociation of H$_2$O, respectively [panel (a)], as well as differences between PECs computed using lower-bound methods (v2RDM and SOS) and the full CI [panel (b)]. All calculations were carried out using the STO-3G basis set. Two sets of v2RDM data are provided, which were performed with and without the imposition of spin-symmetry constraints.\cite{Mazziotti05_052505, Ayers12_014110} For both molecules, at all geometries, both v2RDM (with and without spin-symmetry constraints) and SOS result in lower-bounds to the full CI energy, as expected. For v2RDM, the higher-quality result ({\em i.e.}, the tighter lower-bound to the full CI) is obtained when imposing spin symmetry constraints, as such constraints reduce the space of 2RDMs over which the v2RDM optimization is performed. 
We find better agreement between the full rank-2 SOS and v2RDM in the limit of large interatomic separations, when lifting the spin symmetry constraints in the latter method; which is also expected behavior. 

Figures~\ref{FIG:N2} and \ref{FIG:H2O} also include PECs and errors in lower-bounds associated with more approximate SOS representations of the Hamiltonian built from two subsets of the rank-2 algebra. We consider the SOS built from particle-hole / hole-particle generators plus particle and hole generators [labeled SOS(a$^\dagger$a, aa$^\dagger$, a, a$^\dagger$)] and the spin-free algebra described in Sec.~\ref{SEC:SPIN_FREE_DUAL} [labeled SOS(spin-free)]. \textcolor{black}{These two variations of the level 2 SOS algebra are of the smallest size ($\mathcal{O}(n^{4})$ degrees of freedom) to represent an electronic structure Hamiltonian.} For both molecules, it is clear that eliminating the particle-particle and hole-hole generators from the algebra results in a substantial decrease in the quality of the lower-bound estimate from the SOS. In particular, the error in the lower-bound estimate from the approximate rank-2 algebra is more than two orders of magnitude larger than that from the full rank-2 algebra at the dissociation limit. Even larger errors are observed for the spin-free algebra. We can also see that the shapes of the PECs from the more approximate SOS representations (depicted in the insets of Figs.~\ref{FIG:N2} and \ref{FIG:H2O}) differ substantially from the other curves, particularly in the case of the spin-free algebra. Clearly, the quality of the lower-bound estimates from these two SOS representations is too poor to be of any direct use in chemistry applications. However, even with the spin-free algebra, the SOS structure itself can be useful in quantum computing applications where it can be combined with the spectral gap amplification technique\cite{Boixo13_593} to obtain large speedups in ground-state energy estimation and expectation value estimation with low-energy states~\cite{low2025fast,king2025quantum}.

To give an idea for the degree of potential enhancement in quantum algorithms that would require minimal changes to block encodings constructions~\cite{low2025fast} we compute the SOS representation ($\mathcal{B}_{\text{SOS}}$) for small Iron-Sulfur complexes commonly used as quantum computing compilation benchmarks. We compare the spin-free algebra to a subset of the full level-2 spin-adapted algebra which contains only the quartic elements ${\bf G}$ of Eq.~\eqref{eq:G_same_spin}. The energies and differences from a variational upper bound are tabulated in Table~\ref{tab:system_data} along with the energy gap between the two variational solutions. The square root of the energy gap quantifies the query complexity improvement by including higher algebra components. We tabulate the ratio of the query complexities between spinfree and spinful SOS representations to obtain a proxy for reduction in calls to a block encoding of the SOS. This improvement does not account for the other parameters of quantum circuit complexity and block-encoding normalization~\cite{king2025quantum} for implementing the spinful SOS representation and thus is not a full accounting of the quantum algorithm costs for ground state energy estimation using quantum phase estimation~\cite{low2025fast}. Yet, the results suggest that more careful selection of algebra components could improve quantum algorithm costs further.
\begin{table}[h]
\centering
\scriptsize
\begin{tabular}{|l|r|r|r|l|r|r|r|}
\hline
\textbf{System} & \textbf{Spinfree} & \textbf{spinfulG (G2aa only)} & \textbf{upper bound energy} & \textbf{upper bound type} & \textbf{spinfree gap} & \textbf{spinful G gap} & \textbf{Ratio of sqrts} \\ \hline
Fe$_{2}$S$_{2}$    & -117.582 & -116.887 & -116.6055 & DMRG M=1500 & 0.9764 & 0.281 & 1.861 \\ \hline
Fe$_{4}$S$_{4}$    & -329.079 & -327.931 & -327.2125 & DMRG M=600 & 1.8664 & 0.718 & 1.611 \\ \hline
FeMoco-54 & -272.524 & -270.629 & -269.0602 & DMRG M=500 & 3.4638 & 1.569 & 1.485 \\ \hline
\end{tabular}
\caption{SOS lower bound and variational upper bound energies (in Hartree) for Iron-Sulfur complexes along with improvements to block encoding query complexities considering the algorithm in Ref.~\cite{low2025fast}.}
\label{tab:system_data}
\end{table}

\begin{figure}[htbp]
    \centering
    \includegraphics[width=8.5cm]{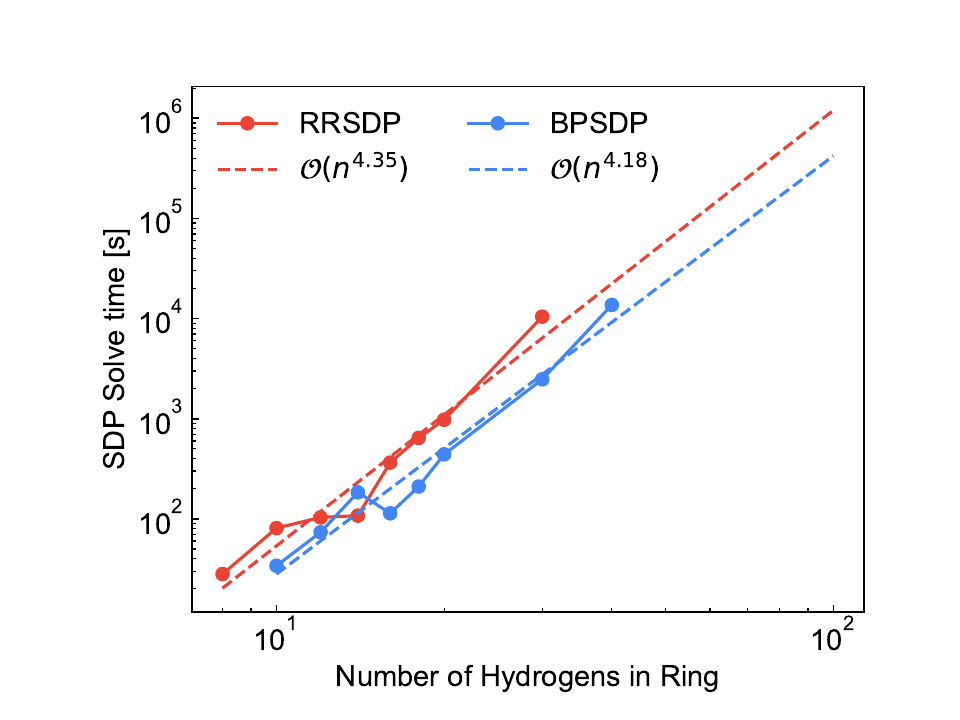}
    \caption{Total semidefinite program solver run time for Hydrogen rings of different size. All calculations use the STO-3G basis and converge the SDP to primal feasibility error and primal-dual gap of $1\times 10^{-5}$. All calculations were performed on an Intel Xeon 2.0 GHz CPU. For both solvers six threads were used for each calculation. The BPSDP complexity is $\mathcal{O}(n^{6})$ for $n$ basis functions coming from eigen decomposing the Gram matrix for the level-2 SOS algebra. The reported scaling does not reflect this as the linear solve is the slowest part in the current libsdp implementation.\label{fig:sdptimings_hrings}}
\end{figure}
Anticipating the use these techniques in a broader simulation context we benchmark spin-free SOS construction time for Hydrogen rings of varying system sizes and extrapolate times to large system to demonstrate the computational feasability of the procedure. In Figure~\ref{fig:sdptimings_hrings} we demonstrate scalings with two SDP solvers for Hydrogen rings of size ten to 30 Hydrogen atoms represented in a minimal basis. The rings are constructed such that the nearest-neighbor Hydrogen atoms are separated by 1.25 \AA. Each SDP is solved to primal L2-norm precision and primal-dual energy gap of $1 \times 10^{-5}$.  The low-rank SDP solver (RRSDP) and boundary-point solver (BPSDP) were implemented in the \texttt{libSDP} library. For RRSDP, the factored solution was set to the full matrix rank. For BPSDP, the frequencey of penalty parameter updates in the outer iteration was set to 3000. We note that increasing this parameter over the standard primal problem (v2RDM) penalty parameter update frequency of 500 was necessary for convergence of the dual SOS SDP. The power-law fits to scalings in Figure~\ref{fig:sdptimings_hrings} indicate that our solvers are limited by the $A\cdot x$ step despite this not being the highest complexity operation. Extrapolated times suggested that the spin-free dual SOS Hamiltonian construction may be feasible within a day of preprocessing for 100 orbital systems.

\section{Conclusion}

We have presented a unified theoretical framework between variational 2-RDM theory and SOS lower bound methods by leveraging the weighted SOS construction~\cite{helton2004positivstellensatz, pironio2010convergent}. The theoretical framework and its connection to various quantum algorithms (BLISS~\cite{loaiza2023block} and SOSSA~\cite{king2025quantum}) are supported by derivations of spin-adapted SOS mathematical programs along with numerics on model systems and Iron-Sulfur complexes. 

Our work highlights that one can use v2RDM or weighted SOS to design bespoke algebras for quantum algorithms and that either framework can be used to generate a Hamiltonian representation that is near-frustration free. To further connect to quantum algorithms or classical algorithms exploiting low-energy assumptions it is clear that a number of improvements must be made. Firstly, the scaling analysis of the SDP program suggests that we are memory speed limited. Further refinement of solvers to avoid the bottlenecks in RRSDP and BPSDP must be devised if we are to scale to large systems. Secondly, strategies for selecting a minimal algebra with efficient quantum implementations~\cite{low2025fast} will be required to balance the cost of lower bound ground-state energy estimates that improve quantum algorithm query complexity and block encoding costs. Finally, studying how this SOS formalism can be extended to different representations of the Hamiltonian, such as first \textcolor{black}{quantized} plane waves, would further broaden the method's impact.

\section{Code Availability}
Software for constructing the global SOS and weighted SOS lower bound SDP can be found at \\ \texttt{https://github.com/ncrubin/sosfermion}.

\vspace{0.5cm}

\bibliography{Journal_Short_Name, deprince, rdm, packages, references}

\appendix

\section{Quartic Dual Mapping Expressions}\label{app:quartic_mapping_conditions}

Recall the form of the rank-2 SOS Hamiltonian defined in Sec.~\ref{SEC:QUARTIC_DUAL}
\begin{align}
\label{EQN:H_SOS_QUARTIC_BLOCKED_APPENDIX}
    \hat{H}_\text{SOS} & = \sum_\gamma O_{G,\gamma}^\dagger O_{G,\gamma} +  \sum_\gamma \sum_{\sigma \neq \tau} O_{G_{\sigma\tau},\gamma}^\dagger O_{G_{\sigma\tau},\gamma} \nonumber \\
    &+ \sum_\sigma \left ( \sum_\gamma O_{D_{\sigma\sigma},\gamma}^\dagger O_{D_{\sigma\sigma},\gamma} + \sum_\gamma O_{Q_{\sigma\sigma},\gamma}^\dagger O_{Q_{\sigma\sigma},\gamma}\right ) \nonumber \\
    &+  \sum_\gamma O_{D_{\alpha\beta},\gamma}^\dagger O_{D_{\alpha\beta},\gamma} + \sum_\gamma O_{Q_{\alpha\beta},\gamma}^\dagger O_{Q_{\alpha\beta},\gamma}
\end{align}
with
\begin{align}\label{EQN:BASIS_APPENDIX}
O_{G,\gamma} &= \sum_{\sigma} \sum_{ij}\left ( g^\gamma_{i\sigma j\sigma}a_{j\sigma}^\dagger a_{i\sigma} + \bar{g}^\gamma_{i\sigma j\sigma}a_{j\sigma}a^\dagger_{i\sigma} \right )\\
O_{G_{\sigma\tau},\gamma} &= \sum_{ij} \left ( g^\gamma_{i \sigma j \tau}a_{j \tau}^\dagger a_{i \sigma} + \bar{g}^\gamma_{i \tau j \sigma}a_{j \sigma}a^\dagger_{i \tau} \right );~~~\sigma \neq \tau \\
O_{D_{\sigma\sigma},\gamma} &= \sum_{i < j} d^\gamma_{i\sigma j\sigma} (a_{j\sigma} a_{i\sigma} - a_{i\sigma} a_{j\sigma} ) \\
O_{Q_{\sigma\sigma},\gamma} &= \sum_{i < j} q^\gamma_{i\sigma j\sigma} (a^\dagger_{j\sigma} a^\dagger_{i\sigma} - a^\dagger_{i\sigma} a^\dagger_{j\sigma} ) \\
O_{D_{\alpha\beta},\gamma} &= \sum_{ij} \left (  d^\gamma_{i\alpha j\beta} a_{j\beta} a_{i\alpha} + d^\gamma_{j\beta i\alpha} a_{i\alpha}a_{j\beta} \right );~~~\sigma \neq \tau \\
O_{Q_{\alpha\beta},\gamma} &= \sum_{ij} \left (  q^\gamma_{i\alpha j\beta} a^\dagger_{j\beta} a^\dagger_{i\alpha} + q^\gamma_{j\beta i\alpha}a^\dagger_{i\alpha}a^\dagger_{j\beta} \right );~~~\sigma \neq \tau
\end{align}
We take the Hermitian square of each generator, sum over the $\gamma$ label, and introduce various positive semidefinite coefficient matrices to obtain
\begin{align}
\label{eqn:Gsos_appendix}
    \sum_\gamma O_{G,\gamma}^\dagger O_{G,\gamma}  &=  \sum_{ijkl} \sum_{\sigma\tau} ( G^{i\sigma j\sigma}_{k\tau l\tau} a^\dagger_{i\sigma}a_{j\sigma}a^\dagger_{l\tau}a_{k\tau} 
    +{G^\prime}^{i\sigma j\sigma}_{k\tau l\tau} a^\dagger_{i\sigma}a_{j\sigma}a_{l\tau}a^\dagger_{k\tau}  \nonumber \\
    &+{G^{\prime\prime}}^{i\sigma j\sigma}_{k\tau l\tau} a_{i\sigma}a^\dagger_{j\sigma}a^\dagger_{l\tau}a_{k\tau} 
    + {G^{\prime\prime\prime}}^{i\sigma j\sigma}_{k\tau l\tau} a_{i\sigma}a^\dagger_{j\sigma}a_{l\tau}a^\dagger_{k\tau} ) \\
    \sum_\gamma  O_{G_{\alpha\beta},\gamma}^\dagger O_{G_{\alpha\beta},\gamma}  &=  \sum_{ijkl} ( G^{i\alpha j\beta}_{k\alpha l\beta} a^\dagger_{i\alpha}a_{j\beta}a^\dagger_{l\beta}a_{k\alpha} 
     + {G^\prime}^{i\alpha j\beta}_{k\beta l\alpha} a^\dagger_{i\alpha}a_{j\beta}a_{l\alpha}a^\dagger_{k\beta} \nonumber \\
     &+  {G^{\prime\prime}}^{i\beta j\alpha}_{k\alpha l\beta} a_{i\beta}a^\dagger_{j\alpha}a^\dagger_{l\beta}a_{k\alpha} 
     + {G^{\prime\prime\prime}}^{i\beta j\alpha}_{k\beta l\alpha} a_{i\beta}a^\dagger_{j\alpha}a_{l\alpha}a^\dagger_{k\beta}) \\
     \sum_\gamma O_{G_{\beta\alpha},\gamma}^\dagger O_{G_{\beta\alpha},\gamma}  &=  \sum_{ijkl} ( {G}^{i\beta j\alpha}_{k\beta l\alpha} a^\dagger_{i\beta}a_{j\alpha}a^\dagger_{l\alpha}a_{k\beta} 
     + {{G}^\prime}^{i\beta j\alpha}_{k\alpha l\beta} a^\dagger_{i\beta}a_{j\alpha}a_{l\beta}a^\dagger_{k\alpha} \nonumber \\
     &+  {{G}^{\prime\prime}}^{i\alpha j\beta}_{k\beta l\alpha} a_{i\alpha}a^\dagger_{j\beta}a^\dagger_{l\alpha}a_{k\beta} 
     + {{G}^{\prime\prime\prime}}^{i\alpha j\beta}_{k\alpha l\beta} a_{i\alpha}a^\dagger_{j\beta}a_{l\beta}a^\dagger_{k\alpha}) \\
     \sum_\gamma O_{D_{\sigma\sigma},\gamma}^\dagger O_{D_{\sigma\sigma},\gamma}  &=  \sum_{i<j, k<l} D^{i\sigma j\sigma}_{k\sigma l\sigma} (a^\dagger_{i\sigma} a^\dagger_{j\sigma} a_{l\sigma} a_{k\sigma} - a^\dagger_{j\sigma} a^\dagger_{i\sigma} a_{l\sigma} a_{k\sigma} - a^\dagger_{i\sigma} a^\dagger_{j\sigma} a_{k\sigma} a_{l\sigma} + a^\dagger_{j\sigma} a^\dagger_{i\sigma} a_{k\sigma} a_{l\sigma} )  \\
     \sum_\gamma O_{Q_{\sigma\sigma},\gamma}^\dagger O_{Q_{\sigma\sigma},\gamma}  &=  \sum_{i<j, k<l} Q_{i\sigma j\sigma}^{k\sigma l\sigma} (a_{k\sigma} a_{l\sigma} a^\dagger_{j\sigma} a^\dagger_{i\sigma} - a_{l\sigma} a_{k\sigma} a^\dagger_{j\sigma} a^\dagger_{i\sigma} - a_{k\sigma} a_{l\sigma} a^\dagger_{i\sigma} a^\dagger_{j\sigma} + a_{l\sigma} a_{k\sigma} a^\dagger_{i\sigma} a^\dagger_{j\sigma} )  \\
     \sum_\gamma O_{D_{\alpha\beta},\gamma}^\dagger O_{D_{\alpha\beta},\gamma} &= \sum_{ijkl} \sum_{\sigma \neq \tau} ( D^{i\sigma j\tau}_{k\sigma l\tau} a^\dagger_{i\sigma}a^\dagger_{j\tau}a_{l\tau}a_{k\sigma} + D^{j\tau i\sigma }_{k\sigma l\tau} a^\dagger_{j\tau}a^\dagger_{i\sigma}a_{l\tau}a_{k\sigma} \nonumber \\
     &+ D^{i\sigma j\tau}_{l\tau k\sigma} a^\dagger_{i\sigma}a^\dagger_{j\tau}a_{k\sigma} a_{l\tau} + D^{j\tau i\sigma}_{l\tau k\sigma} a^\dagger_{j\tau}a^\dagger_{i\sigma}a_{k\sigma} a_{l\tau}) \\
     \label{eqn:Qab_sos_appendix}
     \sum_\gamma O_{Q_{\alpha\beta},\gamma}^\dagger O_{Q_{\alpha\beta},\gamma} &= \sum_{ijkl} \sum_{\sigma \neq \tau} ( Q_{i\sigma j\tau}^{k\sigma l\tau} a_{k\sigma} a_{l\tau} a^\dagger_{j\tau}a^\dagger_{i\sigma} + Q_{j\tau i\sigma }^{k\sigma l\tau} a_{k\sigma} a_{l\tau}a^\dagger_{i\sigma}a^\dagger_{j\tau} \nonumber \\
     &+ Q_{i\sigma j\tau}^{l\tau k\sigma}a_{l\tau}a_{k\sigma}a^\dagger_{j\tau}a^\dagger_{i\sigma}   + Q_{j\tau i\sigma}^{l\tau k\sigma} a_{l\tau}a_{k\sigma}a^\dagger_{i\sigma}a^\dagger_{j\tau})
\end{align}
The coefficient matrices have a block structure given by Eqs.~\ref{eq:G_same_spin} --
\ref{eq:Q_opposite_spin}. Note that this block structure implies that the SOS only includes terms that conserve spin symmetry, {\em i.e.}, those or which the number of $\alpha$-spin (or $\beta$-spin) creation operators equals the number of $\alpha$-spin (or $\beta$-spin) annihilation operators; we have this structure because we limit our considerations to non-relativistic electronic Hamiltonians that preserve spin symmetry.  

Inserting Eqs.~\ref{eqn:Gsos_appendix} -- \ref{eqn:Qab_sos_appendix} into Eq.~\ref{EQN:H_SOS_QUARTIC_BLOCKED_APPENDIX}, bringing each term to a common order, and collecting terms based on operator order gives
\begin{align}
\hat{H}^{(2)}_\text{SOS} = \frac{1}{4} \sum_{\sigma} \sum_{ijkl} a^\dagger_{i\sigma} a_{k\sigma} a^\dagger_{j\sigma} a_{l\sigma} & \left (  G^{i\sigma k\sigma}_{l\sigma j\sigma} \right . - G^{j\sigma k\sigma}_{l\sigma i\sigma}  - G^{i\sigma l\sigma}_{k\sigma j\sigma}  +  G^{j\sigma l\sigma}_{k\sigma i\sigma} \nonumber \\
& - {G^{\prime}}^{i\sigma k\sigma}_{j\sigma l\sigma} + {G^{\prime}}^{j\sigma k\sigma}_{i\sigma l\sigma}  + {G^{\prime}}^{i\sigma l\sigma}_{j\sigma k\sigma}  -  {G^{\prime}}^{j\sigma l\sigma}_{i\sigma k\sigma} \nonumber \\
& - {G^{\prime\prime}}^{k\sigma i\sigma}_{l\sigma j\sigma} + {G^{\prime\prime}}^{k\sigma j\sigma}_{l\sigma i\sigma}  + {G^{\prime\prime}}^{l\sigma i\sigma}_{k\sigma j\sigma}  -  {G^{\prime\prime}}^{l\sigma j\sigma}_{k\sigma i\sigma}  \nonumber \\
& + {G^{\prime\prime\prime}}^{k\sigma i\sigma}_{j\sigma l\sigma} - {G^{\prime\prime\prime}}^{k\sigma j\sigma}_{i\sigma l\sigma}  - {G^{\prime\prime\prime}}^{l\sigma i\sigma}_{j\sigma k\sigma}  + \left . {G^{\prime\prime\prime}}^{l\sigma j\sigma}_{i\sigma k\sigma}  \right ) \nonumber \\
&+ 4 \tilde{D}^{i\sigma j\sigma}_{k\sigma l\sigma} +\left . 4 \tilde{Q}^{k\sigma l\sigma}_{i\sigma j\sigma}  \right ) \nonumber \\
+\frac{1}{2} \sum_{\sigma \neq \tau} \sum_{ijkl} a^\dagger_{i\sigma} a_{k\sigma} a^\dagger_{j\tau} a_{l\tau} & \left (  G^{i\sigma k\sigma}_{l\tau j\tau} \right .  +  G^{j\tau l\tau}_{k\sigma i\sigma} - {G^{\prime}}^{i\sigma k\sigma}_{j\tau l\tau} -  {G^{\prime}}^{j\tau l\tau}_{i\sigma k\sigma} \nonumber \\
& - {G^{\prime\prime}}^{k\sigma i\sigma}_{l\tau j\tau} -  {G^{\prime\prime}}^{l\tau j\tau}_{k\sigma i\sigma} + {G^{\prime\prime\prime}}^{k\sigma i\sigma}_{j\tau l\tau} + {G^{\prime\prime\prime}}^{l\tau j\tau}_{i\sigma k\sigma}  \nonumber \\
& - G^{i\sigma l\tau}_{k\sigma j\tau} +  {G^{\prime}}^{i\sigma l\tau}_{j\tau k\sigma} + {G^{\prime\prime}}^{l\tau i\sigma}_{k\sigma j\tau} - {G^{\prime\prime\prime}}^{l\tau i\sigma}_{j\tau k\sigma}  \nonumber \\
& - G^{j\tau k\sigma}_{l\tau i\sigma} +  {G^{\prime}}^{j\tau k\sigma}_{i\sigma l\tau} + {G^{\prime\prime}}^{k\sigma j\tau}_{l\tau i\sigma} -  {G^{\prime\prime\prime}}^{k\sigma j\tau}_{i\sigma l\tau}  \nonumber \\
&+ D^{i\sigma j\tau}_{k\sigma l\tau} - D^{i\sigma j\tau}_{l\tau k\sigma}- D^{ j\tau i\sigma}_{k\sigma l\tau} + D^{ j\tau i\sigma}_{l\tau k\sigma}  \nonumber \\
&+ Q_{i\sigma j\tau}^{k\sigma l\tau} - Q_{i\sigma j\tau}^{l\tau k\sigma}- Q_{ j\tau i\sigma}^{k\sigma l\tau} + \left . Q_{ j\tau i\sigma}^{l\tau k\sigma}  \right )
\end{align}
\begin{align}
\hat{H}^{(1)}_\text{SOS} = \frac{1}{4} \sum_\sigma \sum_{ij} a^\dagger_{i\sigma} a_{j\sigma} \sum_{p} & \left (  3 G^{i\sigma p\sigma}_{j\sigma p\sigma} \right . + G^{p\sigma p\sigma}_{j\sigma i\sigma} + G^{i\sigma j\sigma}_{p\sigma p\sigma} - G^{p\sigma i\sigma}_{p\sigma j\sigma} \nonumber \\ 
&- {G^{\prime}}^{p\sigma p\sigma}_{i\sigma j\sigma} +3 {G^{\prime}}^{i\sigma j\sigma}_{p\sigma p\sigma} - 3{G^{\prime}}^{i\sigma p\sigma}_{p\sigma j\sigma} + {G^{\prime}}^{p\sigma j\sigma}_{i\sigma p\sigma} \nonumber \\ 
&+ 3 {G^{\prime\prime}}^{p\sigma p\sigma}_{j\sigma i\sigma} -3 {G^{\prime\prime}}^{p\sigma i\sigma}_{j\sigma p\sigma} - {G^{\prime\prime}}^{j\sigma i\sigma}_{p\sigma p\sigma} + {G^{\prime\prime}}^{j\sigma p\sigma}_{p\sigma i\sigma} \nonumber \\ 
&- 3 {G^{\prime\prime\prime}}^{p\sigma p\sigma}_{i\sigma j\sigma} -3 {G^{\prime\prime\prime}}^{j\sigma i\sigma}_{p\sigma p\sigma} + 3{G^{\prime\prime\prime}}^{p\sigma i\sigma}_{p\sigma j\sigma} - {G^{\prime\prime\prime}}^{j\sigma p\sigma}_{i\sigma p\sigma} \nonumber \\ 
&+ 4 \tilde{D}^{i\sigma p\sigma}_{j\sigma p\sigma} - \left . 12 \tilde{Q}^{j\sigma p\sigma}_{i\sigma p\sigma} \right ) \nonumber \\
+ \sum_{\sigma \neq \tau} \sum_{ij} a^\dagger_{i\sigma} a_{j\sigma} \sum_{p}
& \left (  {G^{\prime}}^{i\sigma j\sigma}_{p\tau p\tau} \right . 
+  {G^{\prime\prime}}^{p\tau p\tau}_{j\sigma i\sigma} 
-  {G^{\prime\prime\prime}}^{p\tau p\tau}_{i\sigma j\sigma} -  {G^{\prime\prime\prime}}^{j\sigma i\sigma}_{p\tau p\tau} \nonumber \\ 
&+ G^{i\sigma p\tau}_{j\sigma p\tau} - {G^{\prime}}^{i\sigma p\tau}_{p\tau j\sigma} - {G^{\prime\prime}}^{p\tau i\sigma}_{j\sigma p\tau} +  {G^{\prime\prime\prime}}^{p\tau i\sigma}_{p\tau j\sigma} \nonumber \\ 
&- Q^{j\sigma p\tau}_{i\sigma p\tau} + Q^{p\tau j\sigma}_{i\sigma p\tau} + Q^{j\sigma p\tau}_{p\tau i\sigma} - \left . Q^{p\tau j\sigma}_{p\tau i\sigma} \right ) 
\end{align}
\begin{align}
\lambda = \sum_{\sigma\tau} \sum_{pq} {G^{\prime\prime\prime}}^{p\sigma p\sigma}_{q\tau q\tau} + 2 \sum_\sigma \sum_{pq} \tilde{Q}^{p\sigma q\sigma }_{p\sigma q\sigma } + \sum_{pq} \sum_{\sigma \neq \tau} (Q^{p\sigma q\tau}_{p\sigma q\tau} - Q^{p\sigma q\tau}_{q\tau p\sigma} )
\end{align}
where we have introduced
\begin{align}
\tilde{D}^{i\sigma j\sigma}_{k \sigma l\sigma} & = s_{ij} s_{kl} D^{\text{idx}(i\sigma, j\sigma)}_{\text{idx}(k\sigma, l\sigma)}\\
\tilde{Q}^{k\sigma l\sigma}_{i\sigma j\sigma} & = s_{kl} s_{ij} Q^{\text{idx}(k\sigma, l\sigma)}_{\text{idx}(i\sigma, j\sigma)}
\end{align}
Here,
\begin{align}
s_{ij} = \begin{cases}
1: &i < j \\
-1: &i > j \\
0, &i = j
\end{cases}
\end{align}
and
\begin{align}
\text{idx}(i\sigma, j\sigma) = \begin{cases}
i\sigma j\sigma:  &i < j \\
j\sigma i\sigma: & \text{otherwise}
\end{cases}
\end{align}
Note that these expressions take into account all possible mappings between the SOS and the Hamiltonian, {\em e.g.}, as shown in Eqs.~\ref{eqn:indistinguishability_1} -- \ref{eqn:indistinguishability_4}.
\textcolor{black}{\section{Hubbard Model Lower Bound Proofs}\label{app:hubbard_lower_bounds}
The one-dimensional Hubbard model is 
\begin{align}
H = -t \sum_{\langle i, j\rangle, \sigma}a_{i\sigma}^{\dagger}a_{j\sigma} + U \sum_{i}n_{i\alpha}n_{i\beta}
\end{align}
where $\langle i,j\rangle$ denotes indices $i,j$ that are geometrically nearest-neighbor sites on a lattice, and $\sigma \in \{\alpha, \beta\}$ denotes the spin index of the electrons. Here we derive an analytical lower bound that scales with the size of the lattice and dimensionality of the model. We begin by rewriting the nearest-neighbor hopping Hamiltonian as an SOS. Given 
\begin{align}
O_{ij,\sigma} &= c_{i\sigma} - c_{j\sigma}
\end{align}
then
\begin{align}
O_{ij,\sigma}^{\dagger}O_{ij,\sigma} &= n_{i\sigma} + n_{j\sigma} - \left(c_{i\sigma}^{\dagger}c_{j\sigma} + c_{j\sigma}^{\dagger}c_{i\sigma}\right) \\
\therefore \nonumber \\
t O_{ij,\sigma}^{\dagger}O_{ij,\sigma}  - t\left(n_{i\sigma} + n_{j\sigma}\right) &= -t\left(c_{i\sigma}^{\dagger}c_{j\sigma} + c_{j\sigma}^{\dagger}c_{i\sigma}\right).
\end{align}
Also note that the Hubbard onsite interaction (assuming $U\geq 0$) can be written as sum of squares
\begin{align}
U n_{i\uparrow}n_{i\downarrow} &= U \left(a_{i\uparrow}^{\dagger}a_{i\downarrow}^{\dagger}\right)\left(a_{i\downarrow}^{}a_{i\uparrow}^{}\right).
\end{align}
We can now represent the Hubbard Hamiltonian using the $O_{ij,\sigma}$ generators and the $\sqrt{U}a_{i\downarrow}a_{i\uparrow}$ generators. Let us define the local hopping
\begin{align}
F_{i,\sigma} &= \sum_{j\in \mathcal{N}(i)}O_{ij,\sigma}^{\dagger}O_{ij,\sigma}
\end{align}
such that
\begin{align}
\sum_{i,\sigma}t F_{i,\sigma} &= H_{\text{hubbard-hop}} + 2tz \sum_{i}\left(n_{i\uparrow} + n_{i\downarrow}\right)
\end{align}
where $t$ is the hopping strength (assumed $t\geq 0$), $z$ is the dimensionality of the lattice (in our case $1$), and $H_{\text{hubbard-hop}} $ is the one-body hopping component of the Hubbard model. Combining with the onsite SOS representing the repulsive two-body component we can recover an SOS operator equal to the Hubbard Hamiltonian up to a negative number operator
\begin{align}
\sum_{i\sigma}\left(t F_{i,\sigma} + U\left(a_{i\uparrow}^{\dagger}a_{i\downarrow}^{\dagger}\right)\left(a_{i\downarrow}^{}a_{i\uparrow}^{}\right)\right) - 2tz \hat{N} &= \tilde{H}_{\text{SOS}} - 2tz\hat{N} = H_{\text{hubbard}}.
\end{align}
To convert the left-hand-side of the equation to a full sum of squares we need to convert the $-2tz\hat{N}$ to a non-negative operator and a shift. Employing the anticommutation rules $\{a_{i},a_{j}^{\dagger}\} = \delta_{ij}$ we obtain
\begin{align}
H_{\text{SOS}} &= \sum_{i\sigma}\left(t F_{i,\sigma} + U\left(a_{i\uparrow}^{\dagger}a_{i\downarrow}^{\dagger}\right)\left(a_{i\downarrow}^{}a_{i\uparrow}^{}\right)\right) + 2tz \left(\sum_{i}a_{i\uparrow}a_{i\uparrow}^{\dagger} + a_{i\downarrow}a_{i\downarrow}^{\dagger}\right) \nonumber \\ 
&= H_{\text{SOS}} + \beta\mathbb{1} = H_{\text{hubbard}}
\end{align}
with the lower bound energy certificate $-\beta = -tzL$ for an $L$ site Hubbard model in $z$-dimensions.
\\
\\
We can now demonstrate the Fock-SOS-nn algebra yields $E_{\text{SOS}} = -2Lt$ as well.  Since particle and hole generators are in the nearest-neighbor algebra we use a sum of the $O_{ij\sigma}$ from before and the hole variant
\begin{align}
\tilde{O}_{ij\sigma} &= a_{i\sigma}^{\dagger} + a_{j\sigma}^{\dagger} \\
\tilde{O}_{ij\sigma}^{\dagger}\tilde{O}_{ij\sigma} &= 2 - n_{i\sigma} - n_{j\sigma} - a_{j\sigma}^{\dagger}a_{i\sigma} - a_{i\sigma}^{\dagger}a_{j\sigma}
\end{align}
Summing the local hole SOS with the particle SOS we get
\begin{align}
\tilde{O}_{ij\sigma}^{\dagger}\tilde{O}_{ij\sigma} + O_{ij\sigma}^{\dagger}O_{ij\sigma} = 1 - \left(a_{i\sigma}^{\dagger}a_{j\sigma} + a_{j\sigma}^{\dagger}a_{i\sigma}\right)
\end{align}
multiplying by $t$ and summing over all bonds results in an SOS Hamiltonian where the shift comes from the combination of particle and hole generators resulting in $E_{\text{SOS}} = -2Lt$. For $L=6$ we recover the $-12$ in Figure 2.
\\
\\
Next we derive the lower bound when we add a linear weighting in the SOS. Recall that our multiplier polynomial $f_{j} \in \text{Span}\left(\mathbb{1}, n_{j\alpha}\right)$ and the weighting term is
\begin{align}
\left(f\left(\hat{N} - \eta\right) + \left(\hat{N} - \eta\right)f^{\dagger}\right).
\end{align}
Each $n_{j,\tau}$ subterm of $f$ produces long-range density-density terms
\begin{align}
f_{j}\left(\sum_{i}n_{i\sigma} - \eta\right) = n_{j,\tau}n_{1\sigma} + n_{j\tau}n_{2\sigma} + ... + \eta n_{j\tau}
\end{align}
which must be canceled by something in the SOS part of the weighted sum of squares because the Hamiltonian does not contain long range density-density operators. The restriction to local algebra in Fock-SOS-nn produces $n_{i}$ and $n_{i}n_{i+1}$ like terms and therefore no long-range density-density operators are produced. This forces the $n_{j\tau}$ multipliers in the weighting term to be zero negating the addition to the non-negative ansatz operator. For the identity component of $f$ we return to the discussion of the analytical SOS. Recall,
\begin{align}
\tilde{H}_{\text{SOS}} - 2tz\hat{N} = H
\end{align}
which would imply the weighted SOS must cancel this $-2tz\hat{N}$ term
\begin{align}
H - E_{\text{SOS}}\mathbb{1} &= \text{SOS}_{\text{var}} + c\left(\hat{N} - \eta\right) \\
\tilde{H}_{\text{SOS}} - 2tz\hat{N} - E_{\text{SOS}}\mathbb{1} &= \text{SOS}_{\text{var}} + c\hat{N} - c\eta \label{eq:identity_ideal_term_shift}
\end{align}
where $\text{SOS}_{\text{var}}$ is the local analytical SOS recovering the hopping term.  The equality in Eq.~\eqref{eq:identity_ideal_term_shift} implies $c = -2tz$ and the lower bound constant will be $-\beta = -2t\eta$.
\\
\\
To fully understand the behavior of the linear multiplier, we must also consider the equivalent derivation using the local hole SOS generators. Summing the hole generators over all bonds in the lattice yields an alternative representation of the hopping Hamiltonian:
\begin{align}
\sum_{\langle i,j\rangle, \sigma} t\tilde{O}_{ij\sigma}^{\dagger}\tilde{O}_{ij\sigma} &= 4Lt - 2t\hat{N} + H_{\text{hubbard-hop}} \nonumber \\
H_{\text{hubbard-hop}} &= \sum_{\langle i,j\rangle, \sigma} t\tilde{O}_{ij\sigma}^{\dagger}\tilde{O}_{ij\sigma} + 2t\hat{N} - 4Lt.
\end{align}
Substituting this hole-based representation into the weighted SOS equality, we obtain
\begin{align}
\left(\sum_{\langle i,j\rangle, \sigma} t\tilde{O}_{ij\sigma}^{\dagger}\tilde{O}_{ij\sigma} + 2t\hat{N} - 4Lt\right) - E_{\text{SOS}}\mathbb{1} &= \text{Local SOS} + c\hat{N} - c\eta\mathbb{1}.
\end{align}
To satisfy the operator equality, the SDP matches the local hole SOS and cancels the positive number operator by setting the multiplier to $c = 2t$. Matching the scalar constants then yields
\begin{align}
-4Lt - E_{\text{SOS}} &= -2t\eta \nonumber \\
E_{\text{SOS}} &= 2t\eta - 4Lt.
\end{align}
Because the semidefinite program strictly maximizes the lower bound, it has access to both the particle and hole analytical limits simultaneously and will dynamically select the optimal multiplier $c$ for a given filling $\eta$. The resulting bound is therefore the maximum of these two limits:
\begin{align}
E_{\text{SOS}} &= \max\left(-2t\eta, \; 2t\eta - 4Lt\right).
\end{align}
\\
To resolve the inability of the local algebra to cancel the $n_{i\sigma}n_{j\tau}$ terms generated by the quadratic weighting polynomials, we consider augment the algebra with a global density spanning block
\begin{align}
O_{\text{span}} \in \text{Span}(\mathbb{1}, n_{1\alpha}, n_{1\beta}, \dots, n_{L\beta}).
\end{align}
The Hermitian square of this generator naturally spans all long-range density correlators, allowing the SDP to construct the exact variance penalty operator $\lambda(\hat{N} - \eta)^2$ with $\lambda \geq 0$. This provides the necessary algebraic flexibility for the quadratic multipliers $d_{j\tau}$ to be non-zero. Assuming a uniform quadratic multiplier $d$, the full weighted constraint evaluates to
\begin{align}
W &= 2c(\hat{N} - \eta) + 2d\hat{N}(\hat{N} - \eta).
\end{align}
We can now substitute the analytical particle SOS, the spanning SOS, and the full weighted constraint $W$ into the SDP operator equality:
\begin{align}
\tilde{H}_{\text{SOS}} - 2t\hat{N} - E_{\text{SOS}}\mathbb{1} &= \text{SOS}_{\text{var}} + \lambda(\hat{N} - \eta)^2 + 2c(\hat{N} - \eta) + 2d\hat{N}(\hat{N} - \eta).
\end{align}
Matching the local hopping terms sets $\text{SOS}_{\text{var}} = \tilde{H}_{\text{SOS}}$. We determine the remaining coefficients by expanding the right-hand side and matching powers of the $\hat{N}$ operator. For the $\hat{N}^2$ terms, we have
\begin{align}
0 &= \lambda + 2d \implies 2d = -\lambda.
\end{align}
Matching the linear $\hat{N}$ terms allows the linear multiplier $c$ to simultaneously absorb the cross-terms of the variance penalty and the negative number operator from the Hamiltonian:
\begin{align}
-2t &= -2\lambda\eta + 2c - 2d\eta \nonumber \\
-2t &= -\lambda\eta + 2c \implies 2c = \lambda\eta - 2t.
\end{align}
Finally, matching the scalar constants and substituting the value for $2c$ yields the exact lower bound:
\begin{align}
-E_{\text{SOS}} &= \lambda\eta^2 - 2c\eta \nonumber \\
-E_{\text{SOS}} &= \lambda\eta^2 - \left(\lambda\eta - 2t\right)\eta \nonumber \\
E_{\text{SOS}} &= -2t\eta.
\end{align}
A symmetric matching procedure for the hole SOS similarly yields $E_{\text{SOS}} = 2t\eta - 4Lt$. Through the addition of the global spanning block, the $\lambda\eta^2$ variance terms perfectly cancel in the scalar energy expression. The SDP is therefore free to strictly enforce the targeted particle sector $\eta$ by scaling the penalty $\lambda \to \infty$, successfully lifting the lower bound from the unconstrained floor and fully recovering the analytical V-shaped dispersion but offering no energy improvement.
}
\end{document}